\documentclass[reprint,twocolumn,amsmath,amssymb,superscriptaddress,aps,prd]{revtex4-2}

\usepackage{graphicx}
\usepackage{dcolumn}
\usepackage{bm}
\usepackage[normalem]{ulem}
\usepackage[utf8]{inputenc} 
\usepackage[T1]{fontenc}

\usepackage{aas_macros}
\usepackage{amsfonts}
\usepackage{booktabs}
\usepackage{cancel}
\usepackage{color}
\usepackage{siunitx}
\usepackage{xspace}
\usepackage{float}
\usepackage{url}
   
\usepackage{color} 
\usepackage[colorlinks=true]{hyperref}

\newcommand{\be}{\begin{equation}}
\newcommand{\ee}{\end{equation}}
\newcommand{\bea}{\begin{eqnarray}}
\newcommand{\eea}{\end{eqnarray}}
\newcommand{\lb}{\label}

\newcommand{\dang}{\ensuremath{d_\mathrm{A}}}

\newcommand{\zc}{\ensuremath{z_\mathrm{c}}\xspace}
\newcommand{\ts}{\ensuremath{\theta_\mathrm{s}}\xspace}
\newcommand{\rs}{\ensuremath{r_\mathrm{s}}\xspace}
\newcommand{\rd}{\ensuremath{r_\mathrm{d}}\xspace}
\renewcommand{\aap}{Astron. Astrophys.}
\renewcommand{\mnras}{Mon. Not. R. Astron. Soc.}
\renewcommand{\jcap}{J. Cosmol. Astropart. Phys.}
\renewcommand{\apjl}{Astrophys. J. Lett.}
\renewcommand{\apjs}{Astrophys. J. Suppl.}
\renewcommand{\apj}{Astrophys. J.}
\renewcommand{\aj}{Astron. J.}

\newcommand{\lcdm}{\ensuremath{\Lambda\text{CDM}}\xspace}
\begin{document}


\title[Negative Cosmological Constant]{Negative cosmological constant in the dark sector?}

\author{Rodrigo~Calderón}
 \email{rodrigo.calderon-bruni@umontpellier.fr}
\affiliation{Laboratoire Charles Coulomb, Universit\'e de Montpellier and CNRS}
\author{Radouane~Gannouji}%
 \email{radouane.gannouji@pucv.cl}
\affiliation{Instituto de F\'{\i}sica, Pontificia Universidad Cat\'olica de Valpara\'{\i}so,
Av. Brasil 2950, Valpara\'{\i}so, Chile
}%

\author{Benjamin~L'Huillier}
\email{blhuillier@yonsei.ac.kr}
 \affiliation{Department of Astronomy, Yonsei University, 50 Yonsei-ro, Seodaemun-gu, Seoul 03722, Korea
}%

\author{David~Polarski}
\email{david.polarski@umontpellier.fr}
\affiliation{Laboratoire Charles Coulomb, Universit\'e de Montpellier \& CNRS}

\date{\today}

\begin{abstract}
We consider the possibility that the dark sector of our Universe contains a negative cosmological constant dubbed $\lambda$. For such models to be viable, the dark sector should contain an additional component responsible for the late-time accelerated expansion rate ($X$). We explore the departure of the expansion history of these models from the concordance $\Lambda$ Cold Dark Matter model. For a large class of our models the accelerated expansion is transient with a nontrivial dependence on the model parameters. All models with $w_X>-1$ will eventually contract and we derive an analytical expression for the scale factor $a(t)$ in the neighborhood of its maximal value. We find also the scale factor for models ending in a Big Rip in the regime where dustlike matter density is negligible compared to $\lambda$. We address further the viability of such models, in particular when a high $H_0$ is taken into account. While we find no decisive evidence for a nonzero $\lambda$, the best models are obtained with a phantom behavior on redshifts $z\gtrsim 1$ with a higher evidence for nonzero $\lambda$. An observed value for $h$ substantially higher than $0.70$ would be a decisive test of their viability. 
\end{abstract}

\keywords{Suggested keywords}
\maketitle

\section{Introduction} 
While the physical mechanism behind the late-time accelerated expansion rate of the Universe still remains an open question \cite{2000IJMPD...9..373S,Peebles:2002gy,Padmanabhan:2002ji,Copeland:2006wr}, its phenomenology is known with ever increasing accuracy \cite{Weinberg:2012es,Amendola:2012ys}. It is interesting that perhaps the simplest model, the $\Lambda$ Cold Dark Matter (\lcdm) model, where gravity is described by general relativity whereas dark energy (DE) is simply a positive constant $\Lambda$, can account for the data to some accuracy. Hence the concordance model in which the present accelerated expansion rate is driven by a cosmological constant $\Lambda$ has become the reference cosmological model. Aside from the theoretical problems, the smallness of $\Lambda$ compared to expected Planckian values, it is not clearly established whether this model can successfully cope with all observations especially on small cosmic scales (see e.g. \cite{Bullock:2017xww}). 
Hence one is still investigating other DE models, both inside and outside general relativity, which are able to roughly reproduce the \lcdm phenomenology and are therefore viable with the hope that some specific signature will single them out. 
Recently the so-called tensions with the concordance model, and more generally possible discrepancies \cite{2019PhRvD..99d3506R} between early and late time measurements of cosmic quantities, have attracted a lot of interest with a special emphasis on the $H_0$ tension \cite{Verde:2016ccp,Bernal:2016gxb,2017JCAP...01..015L,2018PhRvD..98h3526S,2019NatAs...3..891V}. 
This latter tension--a substantial difference at the $\sim 4 \sigma$ level between the value of the present Hubble constant $H_0$ derived from the cosmic microwave background (CMB) Planck data \cite{2016A&A...594A..13P} on one hand and from local data on the other hand, when the concordance model is assumed--could imply that the DE sector is more complicated than in the concordance model. This is one more incentive to consider models which are more sophisticated than $\Lambda$CDM. 
It is well known on the other hand that the presence of a positive cosmological constant $\Lambda$ in superstring models is problematic. These theories prefer a negative cosmological constant, dubbed here $\lambda$, reflecting the embedding of the anti de Sitter rather than the de Sitter symmetry group. It is therefore interesting to investigate the possibility that our homogeneous expanding Universe contains a $\lambda$ term and it may come as a surprise that this is indeed viable. In some sense this is so as long as the presence of the $\lambda$ term does not change radically the main properties of the expansion history of our Universe compared to the concordance model. This requires first of all that the (smooth) dark sector (which we call here for simplicity the DE sector) contains an additional component, dubbed here $X$ component, responsible for the late-time accelerated expansion rate (see e.g. \cite{Cardenas:2002np,Visinelli:2019qqu}). 
Note that a transient effective $\lambda$ switching around recombination to a positive $\Lambda$ was considered in \cite{Ye:2020btb} while \cite{Akarsu:2019hmw} considers the intriguing possibility of such a spontaneous switch at $z\sim 2$.
Let us mention that a negative, not necessarily constant, energy component can also appear as a result of the equations of motion like the negative dark radiation component found in \cite{Boisseau:2015hqa}. Among other interesting examples of components that can have negative energy is the ``missing matter'' of \cite{Vazquez:2012ag}, the dynamical $\Lambda(t)$ term in \cite{Grande:2006nn}, or the reconstructed total dark energy component (see e.g. \cite{Dutta:2018vmq}). 

We address in this work the observational viability in the presence of a negative $\lambda$ term for several behaviors of the dark sector, investigating more specifically the constraint coming from a high $H_0$. 
Independently of observations, we study also the future evolution of such universes 
with constant $w_X$, in which case a $\lambda$ term can crucially change the dynamics of our Universe. We address as well the nontrivial appearance of transient accelerated stages in the past. 

\section{Cosmic expansion with a negative cosmological constant}
We recall first the basic equations and concepts.
We intend to study here a universe containing a \emph{negative} cosmological constant $\lambda$. Obviously, such a model cannot accelerate the late-time expansion rate of the Universe in the absence of some additional component in the dark sector.
To comply with observations we add an $X$ component with $w_X<-\frac13$ on very 
low redshifts. For a spatially flat Friedmann-Lema\^itre-Robertson-Walker universe, the evolution of the Hubble parameter as a function of the redshift $z=\frac{a_0}{a}-1$ at $z\ll z_{eq}$ reads 
\be
H^2(z) = H_0^2 \left[ \Omega_{\mathrm{m},0}~(1+z)^3 + \Omega_{\lambda,0} + \Omega_{X,0}~f_X(z) \right],\lb{H2z}
\ee  
where $H(t)\equiv \dot a(t)/a(t)$ is the Hubble parameter, $a$ is the scale factor and a dot stands for the derivative with respect to cosmic time $t$, 
$\Omega_i \equiv \frac{\rho_i}{\rho_{cr}}$ with 
$3H^2\equiv 8\pi G \rho_{cr}$, finally 
$f_X(z)=\frac{\rho_X(z)}{\rho_{X,0}}$ is given by 
\be
f_X(z) = \exp \left[ 3\int_{0}^z dz'~\frac{1+w_{X}(z')}{1+z'}\right], \lb{fz}
\ee
with $w_{X}\equiv p_{X}/\rho_{X}$. 
When we  consider later constraints involving much larger redshifts, we will add radiation and neutrinos. 
It is crucial that in \eqref{H2z} we have 
\be
\Omega_{\lambda,0}<0, \lb{negl}
\ee
as we assume the presence of a negative cosmological constant $\lambda<0$. For such a model it is natural to make the following identification
\be
\Omega_\text{DE,0} \equiv \Omega_{\lambda,0} + \Omega_{X,0} \simeq 1 - \Omega_{\mathrm{m},0}. 
\ee
The combined DE sector must of course be able to produce the late-time accelerated expansion of the Universe.
 
While a negative cosmological constant can hide in the dark sector during the past evolution of our Universe, it can significantly modify its future evolution. 
Before considering the asymptotic future it is also very interesting to study 
the appearance of accelerated stages in such models. 
%
\subsection{Transient accelerated stages}
%
The inclusion of a negative cosmological constant has various interesting features. In this subsection, we study some of its generic properties independent of any observational constraints. 

Let us first remember that for a model of universe composed by matter and dark energy with a constant equation of state (EOS) parameter $w$, the Universe accelerates today for $w<-1/(3\Omega_\text{DE,0})$ which, for example,  gives $w<-0.48$ if $\Omega_\text{DE,0}=0.7$. In the  
presence of two fluids, the problem becomes more complicated. In our case, we consider a negative cosmological constant and a constant equation of state, $w_X$, for the X component. 
The condition of an acceleration of the Universe today reduces to 
\be
w_{X}< \frac{\Omega_{\lambda,0}-1/3}
{1-\Omega_{\mathrm{m},0}-\Omega_{\lambda,0}}.
\ee
We have an additional degree of freedom, $\Omega_{\lambda,0}$.
In fact even if $\Omega_{\lambda,0}+\Omega_{X,0}$ is fixed (to $0.7$ for example), $\Omega_{\lambda,0}$ is free to take any value. Considering $\Omega_{\lambda,0}<0$, we find that the Universe accelerates
today for some value of $w_X$ in the range $-1<w_X<-0.48$ depending on the value of $\Omega_{\lambda,0}$. For a large negative value of $\Omega_{\lambda,0}$, we need a more negative $w_X$, with $w_X\to -1$, in order to produce an acceleration today. Considering now the more complicated situation of an acceleration of the Universe not only today but which could occur at any time, we find some peculiar results when $\Omega_{\lambda,0}<0$. First, it is trivial to see that during the matter era, the two fluids are negligible (assuming $w_X<0$) and therefore the Universe decelerates. In the future, if
the $X$ component is not phantom, the negative cosmological constant eventually dominates and we have recollapse and therefore a nonaccelerating universe. Only if 
the $X$ component is phantom, we will have acceleration in the asymptotic future. Considering the situation where
the $X$ component is not phantom, we have deceleration in the past and in the future. Therefore, we conclude that for nonphantom dark energy with a negative cosmological constant, if the Universe accelerates, it will always be transient.
To have an acceleration, we need the situation where dark energy 
starts to dominate over matter, which always happens at some cosmic time, but it should be sufficiently large at that time in comparison to the negative cosmological constant, otherwise the expansion of the Universe would always be decelerated.
%
\begin{figure}[h]
\includegraphics[scale=0.8]{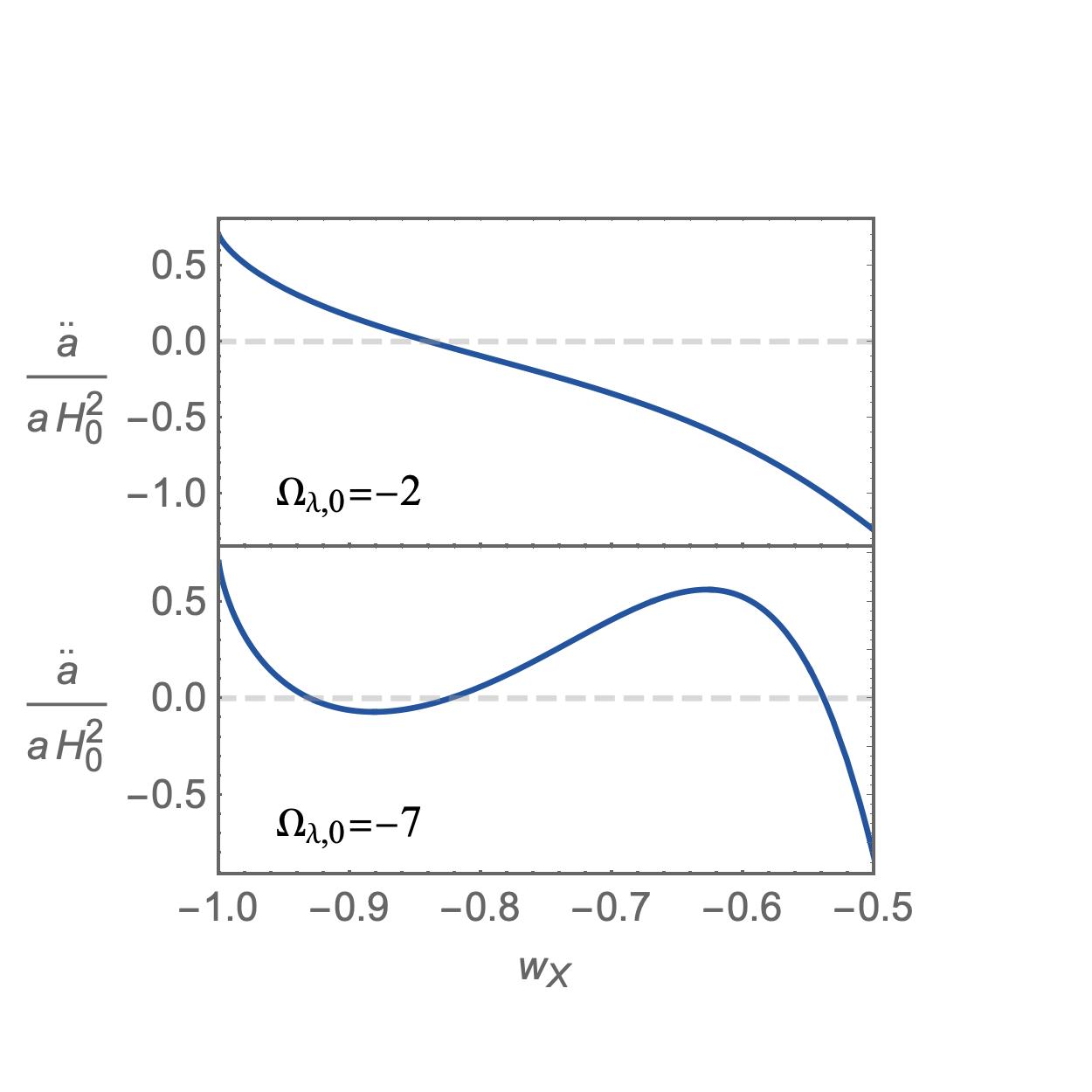}
\caption{Evolution of the maximum value of $\ddot a/a$ reached from matter era until today, as a function of $w_X$ for two different values of $\Omega_{\lambda,0}$ and $\Omega_{\mathrm{m},0}=0.3$.
}
\label{fig:acc}
\end{figure}
%
We see in Fig.~\ref{fig:acc} the maximum of the acceleration reached by the Universe from matter era until today 
%
%
is shown while we assume for illustration $\Omega_{\mathrm{m},0}=0.3$. 
For $\Omega_{\lambda,0}=-2$, we observe that if $-1<w_X<-0.84$, there is an epoch during which we have acceleration while for $-0.84<w_X<-0.5$, the Universe always decelerates until today. This follows our standard intuition: we need $w_X$ sufficiently negative to produce an acceleration. On the other hand, for $\Omega_{\lambda,0}=-7$, we observe a more complicated structure. For $-1<w_X<-0.93$ and $-0.82<w_X<-0.54$ the Universe reached in the past or today a phase of acceleration, while for $-0.93<w_X<-0.82$ and $-0.54<w_X$, the Universe never accelerated in the past. The maximum value for $\ddot a/a$, assuming only $w_X$ constant, is given by
\begin{multline}
\frac{\ddot a}{H_0^2 a}\bigg\rvert_{\rm max}=
 \Omega_{\lambda,0}\\
 -\frac{w_X\Omega_{\mathrm{m},0}}{2(1+w_X)}\left[\frac{\Omega_{X,0}(1+3w_X)(1+w_X)}{-\Omega_{\mathrm{m},0}}\right]^{-\frac{1}{w_X}},
\end{multline}
which shows the nontrivial dependence of the acceleration parameter on $w_X$ and $\Omega_{\lambda,0}$.
In Fig.~\ref{fig:trans}, we extended this analysis to a large range of $\Omega_{\lambda,0}$ and $w_X$ and assuming $\Omega_{\mathrm{m},0}=0.3$. In white, the Universe accelerates today while in gray the Universe decelerates today. The latter is divided in areas (light gray) where the Universe never accelerated until today and situations (darker gray) where the Universe had an acceleration in the past but does not accelerate today.

{Having in mind the observational constraints which will be addressed more thoroughly later, we have also represented in the same figure the set of parameters satisfying Eq.~\eqref{constraint} in red for $h=0.74$ and in purple for $h=0.72$.}
Notice that if $\Omega_{\lambda,0}=0$, we need $w_X=-1.2$ to obtain $h=0.74$ as it was already noticed in \cite{Alestas:2020mvb}.
For negative values of the cosmological constant the required phantomness is milder as we will see in Sec.~III.

\begin{figure}[ht] 
\includegraphics[width=\columnwidth]{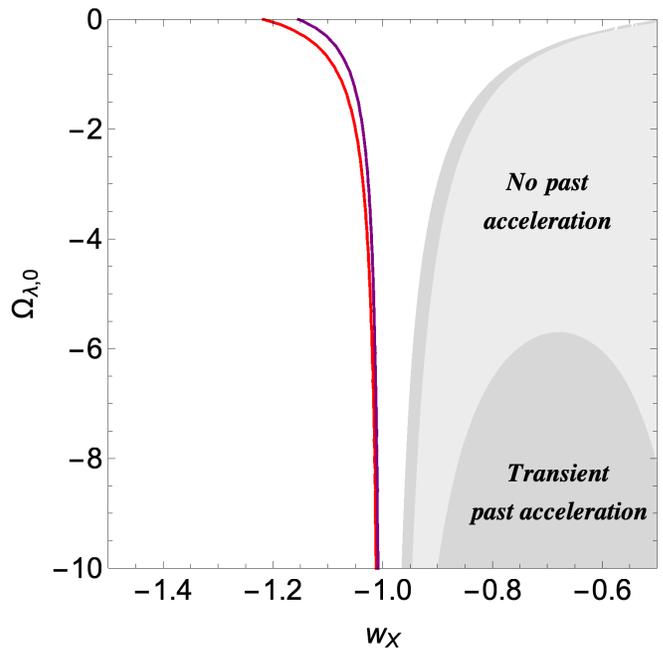}
\caption{Evolution of the Universe in the space $(w_X,\Omega_{\lambda,0})$. In white, the Universe accelerates today, in light gray, the Universe never accelerated until today while in darker gray, we have a transient situation where the Universe accelerated in the past but does not accelerate today. 
These results do not rely on any observational constraints, only 
$\Omega_{\mathrm{m},0}=0.3$ and $w_X$ constant are assumed. 
For comparison, we have added in red and purple lines the $(\Omega_{\lambda,0},w_X)$ values which satisfy Eq.~\eqref{constraint} for $h=0.74$ and $h=0.72$ respectively.}
\label{fig:trans}
\end{figure}

%
\subsection{Future asymptotic solutions}
Let us consider now the limit in the future where dustlike matter density becomes negligible compared to the dark sector density. As we have our Universe in mind, we consider the regime in the future for which 
$\Omega_{\mathrm{m},0} \left( \frac{a_0}{a} \right)^3\ll \left|\Omega_{\lambda,0}\right|< \Omega_{X,0}$. 
Then, we have to solve the effective Friedmann equation 
\be
H^2 = -|\alpha| + \frac{\beta}{a^p}, \lb{Fr1}
\ee
where we have set for brevity 
\bea
\alpha &\equiv& \Omega_{\lambda,0} ~H_0^2 < 0 \nonumber \\ 
\beta &\equiv& \Omega_{X,0} ~H_0^2 ~a_0^p \nonumber \\ 
p &\equiv& 3(1+w_X). 
\eea
The exact solution of Eq.~\eqref{Fr1} is 
\be
a = \left( \frac{\beta}{|\alpha|} \right)^{\frac{1}{p}} \sin^{\frac{2}{p}} 
                      \left( \frac{p}{2} \sqrt{|\alpha|}~t + C \right), \lb{a}
\ee
or more explicitly in function of the cosmological parameters
\be
\frac{a}{a_0} = \left( \frac{\Omega_{X,0}}{\left|\Omega_{\lambda,0}\right|} \right)^{\frac{1}{p}}
   \sin^{\frac{2}{p}} \left( \frac{p}{2} \sqrt{\left|\Omega_{\lambda,0}\right|} ~H_0 t + C \right),
\lb{a1}
\ee 
where $C$ is an integration constant. We obtain further
\bea
\frac{H}{H_0} &=& \sqrt{\left|\Omega_{\lambda,0}\right|} ~\cot \left(\frac{p}{2} 
\sqrt{\left|\Omega_{\lambda,0}\right|} ~H_0 t + C  \right) \lb{H} \\
\frac{{\dot H}}{H_0^2} &=& \frac{p}{2} \left|\Omega_{\lambda,0}\right| \left[
   -1 - \cot^{2} \left( \frac{p}{2} \sqrt{\left|\Omega_{\lambda,0}\right|} ~H_0 t + C  \right)
                                               \lb{dotH}  \right]
\eea
If $p=0~(w_X=-1)$, the two terms on the rhs of 
Eq.~\eqref{Fr1} combine to give an effective positive cosmological constant. The resulting 
future evolution is that of $\Lambda$CDM. Note that it is possible in this case to give the exact analytic expression even when dustlike matter is taken into account. 

We consider next $p>0$; in other words the X component is not of the phantom type. As its 
density decreases with expansion, the Universe will eventually recollapse. Note that 
the density of dustlike matter decreases even more rapidly ($\propto a^{-3}$) so 
Eq.~\eqref{Fr1} applies if $\left|\Omega_{\lambda,0}\right|\ll \Omega_\text{DE,0}\approx 0.7$ 
and $w_X\approx -1$. Indeed, we can read off from Eq.~\eqref{a1} the condition for the existence of a time interval before the contraction during which dustlike matter can be neglected, viz. 
\be
\left( \frac{ \left|\Omega_{\lambda,0}\right| }{ \left|\Omega_{\lambda,0}\right| + \Omega_\text{DE,0} } 
    \right)^{\frac{3}{p}} \ll \frac{ \left|\Omega_{\lambda,0}\right| }{ \Omega_{\mathrm{m},0} } \lb{cond}
\ee
As expected, we verify further with Eqs.~\eqref{H} and \eqref{dotH} that the 
expansion is decelerating, ${\ddot a <0}$, (at least) in the neighborhood of $a_M$, the 
maximal value of the scale factor. As expected this is only so for $0<p<2$ ($-1<w_X<-\frac13$); 
otherwise there is no acceleration at all.  
So even if the Universe is accelerating today, it passes again through $\ddot a =0$, from an 
accelerating to a decelerating expansion rate. When \eqref{cond} is satisfied, this takes 
place at 
\be
\frac{a_M}{a}\simeq \left( 1+\frac{p}{2} \right)^{\frac{1}{p}}, 
\ee
which is close to $a_M$ and lies in the regime described by \eqref{Fr1}.

We now turn to $p<0$, the phantom case. It is clear that the Universe will eventually reach 
the Big Rip singularity in a finite time $t_{\infty}$. In some range before $t_{\infty}$, the 
dustlike component will be negligible compared to the negative cosmological constant. In that 
case, $a(t)$ is given to high accuracy by the solution \eqref{a} or \eqref{a1}. To ensure the 
presence of a Big Rip at $t_{\infty}$, we write the integration constant $C$ in a way to have 
a Big Rip at $t=t_{\infty}$ and the solution for $a(t)$ then reads
\be
a(t) = \left( \frac{\left|\Omega_{\lambda,0}\right|}{\Omega_{X,0}} \right)^{\frac{1}{|p|}} a_0 
  \sin^{-\frac{2}{|p|}} 
   \left( \frac{|p|}{2} \sqrt{\left|\Omega_{\lambda,0}\right|} ~H_0(t_{\infty}-t) \right).\lb{aBR}
\ee 
We verify easily that in the limit $t\to t_{\infty}$, this solution tends asymptotically to 
\be
a(t) \sim \frac{A}{ (t_{\infty} - t)^{\frac{2}{|p|}} },
\ee
with $A$ given by 
\be
A = \left( \frac{|p|}{2} \sqrt{\Omega_{X,0}} ~H_0 \right)^{-\frac{2}{|p|}} a_0.  \lb{aBRas}
\ee
This is the well-known singular behavior in the vicinity of $t_{\infty}$, depending solely 
on the phantom component, here the $X$ component. The solution \eqref{aBR} gives a nearly 
exact fit in the regime $\Omega_{\mathrm{m},0}\ll \left|\Omega_{\lambda,0}\right|$ and improves on 
\eqref{aBRas} when $t$ is sufficiently far from $t_{\infty}$. 

\section{Cosmic Relevance of $\lambda$ and the \texorpdfstring{$H_0$}{H0} tension}
\label{sec:h0}
We now turn to the cosmic relevance of models admitting a negative cosmological constant $\lambda$. As for any cosmological model differing from \lcdm, an important question to address is how viable the model is if the measured value of $h$ is substantially higher than $0.67$. 
It is well known that there is a tension between the value of $H_0$ obtained by Planck and the value obtained with many local (low redshifts) measurements. 
This is a very interesting problem which has been widely investigated recently in various ways (see e.g. \cite{Colgain:2018wgk,Poulin:2018cxd,Vattis:2019efj,Agrawal:2019dlm,2019ApJ...883L...3L,Vagnozzi:2019ezj,2020PDU....3000666D,Demianski:2019vzl,Anchordoqui:2019amx,2020PhRvD.101j3517B,Hernandez-Almada:2020uyr,Barker:2020gcp,2020PhRvL.125r1302J,2020JCAP...10..044B,Banerjee:2020xcn,Sola:2020lba,2020arXiv200703381S,Benaoum:2020qsi} for a nonexhaustive list). 
This tension can be traced back to the measurement of the standard ruler $\rs$, the comoving sound horizon at   recombination time (very close to the drag epoch) relevant for the corresponding angle $\ts$ sustended on the CMB 
\be\lb{rs}
\rs(z_1) = \int_0^{t_1} c_\mathrm{s} \frac{dt}{a(t)} = \frac{1}{a_0} \int_{z_1}^{\infty} c_\mathrm{s} \frac{dz}{H(z)},
\ee 
where adiabatic primordial fluctuations are assumed. The angle $\ts$ is given by 
\be
\ts = \frac{a_1 ~\rs(z_1)}{d_\mathrm{A}(z_1)}, \lb{thetas1}
\ee
where $d_\mathrm{A}(z)$ is the angular-diameter distance out to a redshift $z$. We finally obtain 
\be
\ts = \frac{\rs(z_1)}{r(z_1)}, \lb{thetas}
\ee
with 
\be
r(z_1) = \frac{c}{a_0} \int_0^{z_1} \frac{dz}{H(z)}. \lb{r}
\ee
We have written explicitly the light velocity $c$ and $a_0$ in \eqref{rs},  \eqref{r} ($c=1$ and $a_0=1$ in this work but will sometimes be written explicitly). 
We choose the Planck 2018 TT,TE,EE+LowE+Lensing constraints [no baryon acoustic oscillations (BAO)], with one massive neutrino species \cite{2020A&A...641A...6P}.
We take the redshift $z_1=z_{\rm rec}=1089.92, ~\ts^{\rm Planck} = 1.04110 \times 10^{-2},~\rs^{\rm Planck}(z_{\rm rec})=\SI{144.43} {Mpc}$; see Table \ref{tab:fix}. 
 
The relative energy density $\Omega_{i,0}\equiv \frac{\rho_{i,0}}{\rho_\mathrm{cr,0}}$, defined as 
\be
\Omega_{i,0} = \frac{8\pi G \rho_{i,0}}{3 H_0^2},
\ee
suffers from the uncertainty of the value of $H_0$ even if $\rho_i$ is otherwise known. However, it is often possible to find observationally the numerical value of the combined quantities
\be
\omega_i\equiv \Omega_{i,0}~h^2, 
\ee
with $h\equiv \frac{H_0}{\SI{100}{km.s^{-1}.Mpc^{-1}}}$. For our models, we have obviously at low redshifts 
\begin{multline}
H(z) = \left[ \omega_m ~(1+z)^3 + \omega_{\lambda} + \omega_X~f_X(z) \right]^{\frac12}\\ 
\times \SI{100}{km.s^{-1}.Mpc^{-1}}. \lb{Homega}
\end{multline}
In the standard $\Lambda$CDM model, the value of $\rs(z_1)$ is controlled by the quantities $\omega_i$ contained in the model with the notable exception of $\omega_{\Lambda}$. Measuring these quantities yields in turn $\rs^{\rm Planck}(z_1)$. As the angle $\ts^{\rm Planck}$ is accurately measured by the Planck Collaboration, the numerical value of $r(z_1)$ becomes 
fixed in turn (for given $\ts$ measured by Planck) to its value $r^{\rm Planck}(z_1)$, 
\be
r^{\rm Planck}(z_1) =\SI{13872.8}{Mpc}. \label{rplanck}
\ee
Hence, once a cosmological model is adopted which does not change the early time physics of $\Lambda$CDM, such a model is compelled to give the same $r(z_1)$. For $\Lambda$CDM this boils down to fix the value $\omega_{\Lambda}^{\rm Planck}$ and therefore the value of $H_0$. 
The Planck collaboration finds $H_0=\SI{67.36}{km.s^{-1}.Mpc^{-1}}$ a value substantially lower than the value measured locally. We refer the interested reader to the excellent account given in \cite{Knox:2019rjx}. 

In this work we are interested in models which depart from $\Lambda$CDM regarding the Universe expansion for $z<z_1$; however in a way they satisfy  
\be
r(z_1) = r^{\rm Planck}(z_1) \label{constraint}
\ee
for a \emph{larger} $H_0$ value. 
At this stage we emphasize two points: first, as we have said earlier, when we compare our models with data we fix $\rs$, and hence we must also fix $\omega_m$ and $\omega_r$, to their fiducial Planck values. 
A more elaborate analysis could be performed by allowing all $\omega_i$'s to vary. We feel the constraint on our models is essentially expressed with our simpler analysis at this stage.  
Second, in our theoretical investigations performed below in Sec.~III (see Figs.~\ref{Fig3} and~\ref{Fig4}), the constraint \eqref{constraint} is used, assuming that both $\rs$ \emph{and} $\ts$ are fixed to their Planck values (see Table \ref{tab:fix}). 
In the comparison with observations performed in Sec.~IV, 
$\rs$, $\omega_m$, and $\omega_r$ are fixed and we consider the measured $\ts$ 
with its associated error; hence $r(z_1)$ can vary around the value \eqref{rplanck}. We can then compare $\theta_\text{s}^\text{theory}$ = $\rs / r(z_1)$ with the Planck measured value $\simeq\num{1.04e-2}$, constraining $\omega_X,~\omega_{\lambda}$ and ultimately $h$. 

It is obvious from \eqref{Homega}, \eqref{thetas} that a larger $H_0$ requires 
\be
\omega_\text{DE} = \omega_{\lambda} + \omega_X > \omega_{\Lambda}^{\rm Planck},
\ee
and a phantom behavior of the $X$ component, hence also an effective phantom behavior of the DE sector. This amounts to explore models with $w_X<-1$ at least during part of the late-time expansion.
It is straightforward to obtain the exact equality 
\be
\omega_\text{DE} - \omega_{\Lambda}^{\rm Planck} = h^2 - (h^{\rm Planck})^2.
\ee
It is further clear that our models satisfy (by construction)
\bea
\Omega_{\mathrm{m},0} &=& \left( \frac{h^{\rm Planck}}{h}\right)^2 \Omega_{\mathrm{m},0}^{\rm Planck} < 
                                                           \Omega_{\mathrm{m},0}^{\rm Planck}\\
\Omega_\text{DE,0} &\simeq& 1 - \Omega_{\mathrm{m},0} > \Omega_{\Lambda,0}^{\rm Planck}.
\eea 
Since we have to calculate distances up to $z=z_{\rm rec}$, where radiation is subdominant but not negligible, we have to properly take into account the effect of photons and neutrinos.
At high redshifts ($z\gtrsim 50$), Eq.~\eqref{Homega} becomes~\cite{2011ApJS..192...18K}
\begin{align}
 &\frac{H(z)}{\SI{100}{km.s^{-1}.Mpc^{-1}}}
  = \Bigg[ \omega_m ~(1+z)^3 + \omega_{\lambda} + \omega_X~f_X(z) ~+ ~~~~~\nonumber\\
 & + \omega_{\gamma}~(1+z)^4 \left( 1+0.2271\,\frac{N_\text{eff}}{3}\, \sum_i f_\nu \left(\frac{m_{\nu_i}}{T_{\nu}}\right) \right) ~\Bigg]^{\frac12},  \lb{Hnu} 
 \end{align}  
with $\omega_{\gamma}=2.47\times 10^{-5}$ while $f_\nu$ is well fitted with $f_\nu(y) \simeq (1+(Ay)^p)^{1/p}$ with $A=\frac{180 \zeta(3)} {7\pi^4}$ and $p = 1.83$ \cite{2011ApJS..192...18K}.
The function $f_\nu$ interpolates between the relativistic behavior, $m_\nu \ll T_\nu$ ($T_\nu\sim a^{-1}$), and the nonrelativistic regime, $m_\nu \gg T_\nu$.
Even for $m_\nu$ as light as $0.06 \si{eV}$, the transition occurs rather early around $z\simeq 110$. Given these considerations, we fix the early Universe cosmology as in Table~\ref{tab:fix}. 
\begin{table}[h!]
    \centering
    \begin{tabular}{ccccccc}
    \toprule
        $100\, \omega_\text{b}$ & $100\, \omega_\text{c}$ & $N_\text{eff}$ & $m_\nu$& $r_\mathrm{s}$  & $r_\mathrm{d}$ & $100\,\ts$ \\
        &&& (\si{eV}) &  (\si{Mpc}) & (\si{Mpc}) &\\
    \midrule
      2.237 & 12.00 &   3.046 & (0,0,0.06) & 144.43 & 147.09 & ${1.04110}$ \\
      \bottomrule
    \end{tabular}
    \caption{Parameter values as given by the Planck 2018 TT,TE,EE+LowE+lensing results (Table~II). $\rs$ is the comoving sound horizon at recombination ($z_1=1089.92$), and \rd at the drag epoch $z_\mathrm{d}=1059.94$}
    \label{tab:fix}
\end{table}

\subsection{Models}
When assessing the viability of our models with respect to the low-redshift data and the constraint on their free parameters coming from  a high $H_0$, we will consider various types of EOS parameters $w_X$ and various values of $\omega_{\lambda}$. 

\vspace{10pt}
\noindent
{Scenario} $\bm w$:
We consider first models with constant equation of state $w_X=$ constant.
In this case we obviously have 
\be
f_X(z) = (1+z)^{3(1+w_X)}.
\ee
As we have seen earlier, to ease the $H_0$ tension necessarily requires a phantom behavior 
and for a constant $w_X$ the only possible choice is to take $w_X<-1$.
We take an EOS of the form
\be
w_X = -1 + \Delta_1 = w_0,~~~~~~~~~~~~~~~~~~\Delta_1<0,
\ee
and we obtain immediately
\be
f_X(z) = (1+z)^{3\Delta_1}.
\ee
While a constant $w_X$ gives us some insight, it is clearly advisable to explore also models with varying equations of state. 

\vspace{10pt}
\noindent
{Chevallier-Polarski-Linder (CPL) scenarios}: 
Here we adopt the CPL parametrization of $w_X$ corresponding to a smoothly 
(differentiable) varying EOS with 
\be
w_X = w_0 + w_a (1-a) \equiv -1 + \Delta + w_a (1-a)
\ee
which gives \cite{Chevallier:2000qy},\cite{Linder:2002et}
\be
f_X(z) = (1+z)^{3(\Delta + w_a)}~e^{-3w_a\frac{z}{1+z}}.
\ee
We consider also two constrained versions:  
   CPL$w_a$, with $w_0=-1$ while $w_a$ is free;
    and CPL$w_0$, where  $w_0$ is free and $w_a=-1-w_0$ so that $w_X\to -1\equiv w_{\infty}$.

\vspace{15pt}
\noindent
{Scenario I}: 
In this scenario, we take a piecewise constant $w_X$ where dark energy is of a phantom type below some transition redshift $z_c$ and a cosmological constant $\Lambda$ above, 
with
\begin{equation}
     w_{X}(z)=
    \begin{cases}
      -1+\Delta_1 = w_0, & \text{for}\ z\le \zc~~~~~~~~\Delta_1<0\\
      -1 = w_{\infty}, & \text{for}\ z>\zc.
    \end{cases}
    \lb{w1}
\end{equation}
Here we take $\Delta_1<0$ in order to ensure a phantom behavior and we fix $z_c=1$. As $z_c$ increases, it is easier to meet the data on small redshifts but it requires stronger phantomness on large redshifts 
in order to comply with the observed $\ts$ and a higher $H_0$. We note that most of the SN Ia data are in the range $z_c\leq 1$. 
In this scenario, the evolution of the X component is given by

\begin{equation}
     f_X(z)=
    \begin{cases}
      (1+z)^{3\Delta_1}, & \text{for}\ z\le \zc \\
      (1+z_c)^{3\Delta_1}, & \text{for}\ z>\zc.
    \end{cases}
    \lb{f1}
\end{equation}

\vspace{10pt}
\noindent
{Scenario II}: 
This scenario has also a piece-wise constant $w_X$, but it is now opposite to the previous scenario, that is 
    \begin{equation}
     w_X(z)=
    \begin{cases}
      -1 = w_0, & \text{for}\ z\le \zc \\
      -1+\Delta_2 = w_\infty, & \text{for}\ z>\zc~~~~~~~~~~~\Delta_2<0
    \end{cases}
    \lb{w2}
    \end{equation}
We have now
\begin{equation}
     f_X(z)=
    \begin{cases}
      1, & \text{for}\ z\le \zc\\      
      \left( \frac{1+z}{1+z_c} \right)^{3\Delta_2}, & \text{for}\ z>\zc. 

    \end{cases}
    \lb{f2}
    \end{equation}
  In this case too, we take $z_c=1$. 

\vspace{10pt}
\noindent
{Scenario III:}
    \begin{equation}
      w_{X}(z)=
     \begin{cases}
         -1+\Delta_1 = w_0, & \text{for}\ z\le z_{c1}~~~~~~~~~~~\Delta_1<0\\
         -1, & \text{for}\ z_{c1}<z\le z_{c2} \\
         -1+\Delta_2 = w_{\infty}, & \text{for}\ z>z_{c2}~~~~~~~~~~~\Delta_2<0.
     \end{cases}
     \lb{w3}
     \end{equation}
 We have in this case
 \begin{equation}
      f_X(z)=
     \begin{cases}
       (1+z)^{3\Delta_1}, & \text{for}\ z\le z_{c1} \\
       (1+z_{c1})^{3\Delta_1}, & \text{for}\ z_{c1}<z\le z_{c2}\\
       \frac{ (1+z_{c1})^{3\Delta_1} }{ (1+z_{c2})^{3\Delta_2} } (1+z)^{3\Delta_2}, 
                         & \text{for}\ z> z_{c2}
     \end{cases}
     \lb{f3}
     \end{equation}
For this scenario we take $z_{c1}=0.1$ and $z_{c2}=1$. A significant change in $w_X$ on very small redshifts is viable and we exploit also this possibility here. 

\begin{figure}[t]
\includegraphics[width=.37\textwidth]{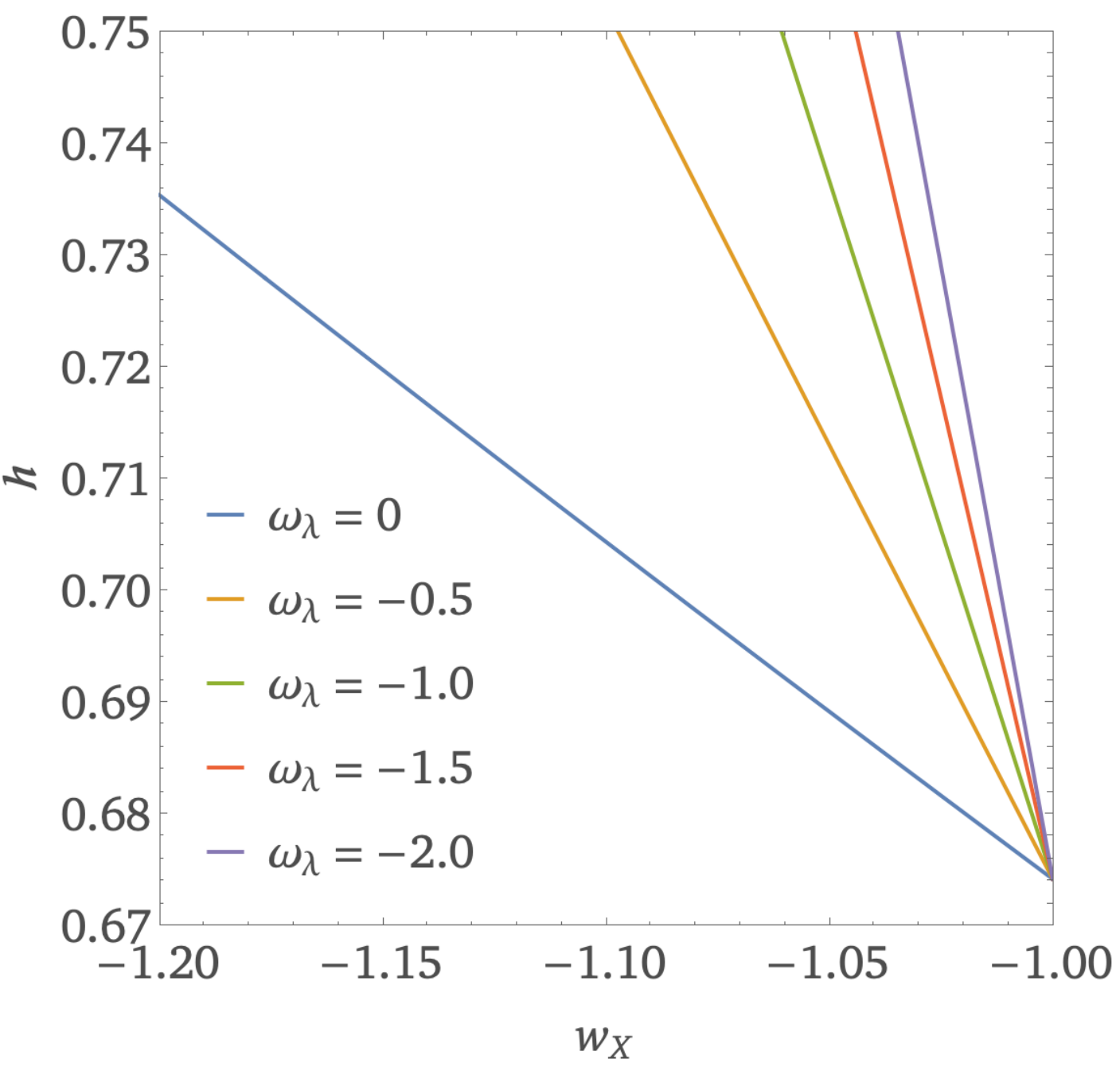}
\includegraphics[width=.4\textwidth]{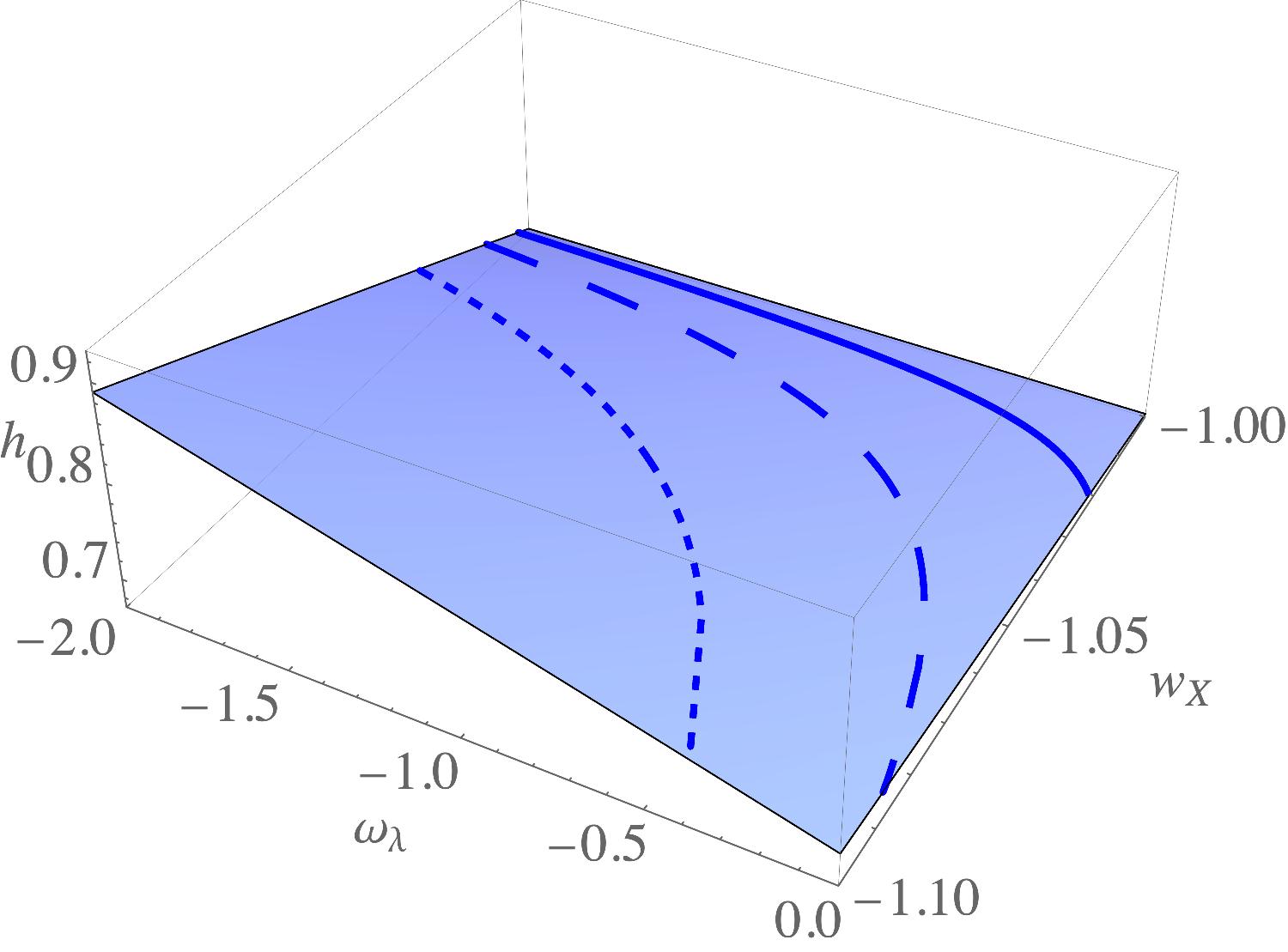}
\caption{Parameters $(h,\omega_\lambda,w_X)$ which satisfy the relation (\ref{constraint}) for models with constant $w_X$. (a) In the upper panel, $\omega_\lambda$ is fixed and we see that an increasing $\left|\omega_{\lambda}\right|$ (from left to right) gives the same $h$ with less phantomness. We note the quasilinear relation, which follows from \eqref{h1} for constant $\omega_\lambda$.  (b) In the lower panel, all parameters are free. For the latter, we show the lines corresponding to a constant $h$ when \eqref{constraint} is satisfied: continuous line ($h=0.68$), long dashed line ($h=0.7$), and dashed line ($h=0.74$). They satisfy to good accuracy \eqref{h1}.} 
\label{Fig3}
\end{figure}

\vspace{10pt}
\par
Once a specific model is adopted we can find the dependence of $h$ on the model parameters for a model satisfying \eqref{constraint}. This gives insight into the phenomenology of these models irrespective of the observational constraints. 

For constant $w_X$, we can study its behavior in terms of the following three free parameters: $\omega_{\lambda}$, $w_X$ and $h$. Indeed the parameter $\omega_X$ is fixed once $\omega_{\lambda}$ and $h$ are given. 
From the constraint \eqref{constraint} however, only two free parameters are left. 
In Fig.~\ref{Fig3}, we show the value of $h$ in terms of $w_X$, when $\omega_{\lambda}$ is fixed and as a function of $w_X$ and $\omega_{\lambda}$ in a 3-dimensional plot. 
In the first case, we confirm the linear relation obtained in \cite{Alestas:2020mvb} (for $\omega_\lambda=0$), which we have generalized here to
\begin{align}
    h = 0.673 + (w_X+1) (0.93 \omega_\lambda-0.33).\lb{h1}
\end{align}

\begin{figure*}[t]
\begin{centering}
\includegraphics[width=1\textwidth]{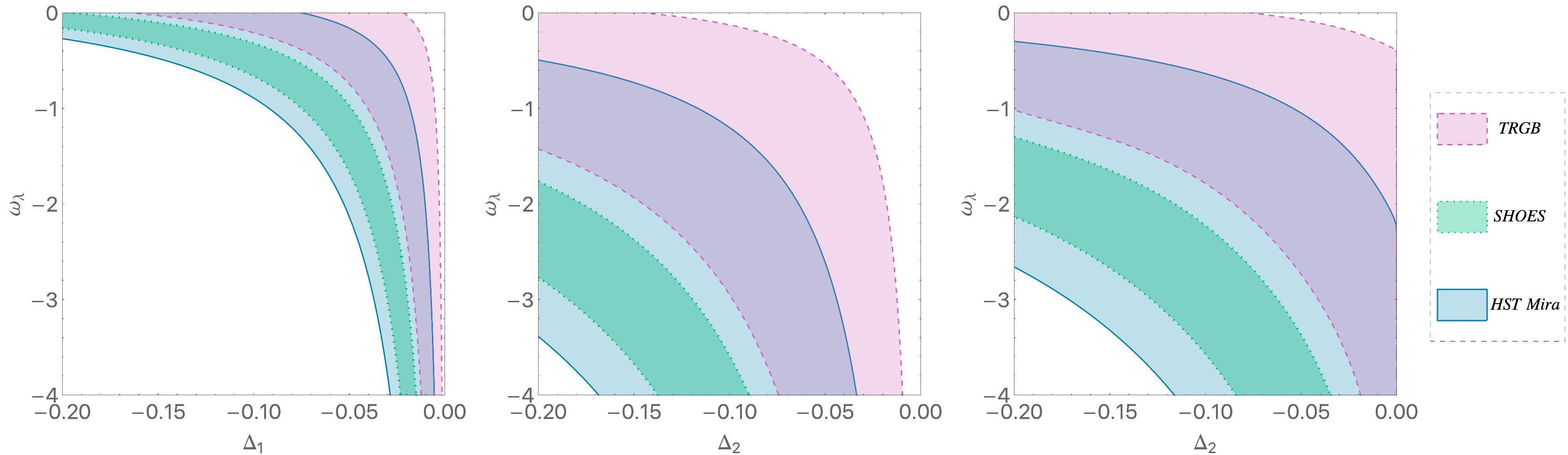}
\par\end{centering}
\caption{We show iso-$h$ curves for scenarios I (left), II (middle), and III with $\Delta_1=-0.04$ (right-hand panel) in the $(\Delta_{1,2},\omega_{\lambda})$ parameter plane. The shaded areas correspond to the following $H_0$ values (in \si{km.s^{-1}.Mpc^{-1}} units): $73.30\pm 4$ (HST-Mira) \cite{2020ApJ...889....5H}), $74.03\pm 1.42$ (SH0ES) \cite{Riess:2019cxk} and $69.8\pm 1.9$ (TRGB) \cite{2019ApJ...882...34F}. }  
\label{Fig4}
\end{figure*}

At this point, we emphasize another interesting aspect of our models which we discuss here for a constant $w_X$. As we will see later, observations favor models of the phantom type, $w_X<-1$. This implies that $\rho_\text{DE}=\rho_{\lambda}+\rho_X$ will necessarily become negative in the past at some redshift $z_{\lambda}$ and it is straightforward to find 
\be
1 + z_{\lambda} = \left[ 1 + \frac{ h^2 - \omega_m }
{ \left|\omega_{\lambda}\right| } 
                 \right]^{-\frac{1}{3(1+w_X)}}. \lb{zl}
\ee
We have used $\omega_\text{DE}\approx h^2 - \omega_m$ which is valid to high accuracy. On the other hand, $H^2(z)$ is necessarily positive $\forall z>z_m$ with 
\be
1+z_m = \left[ \frac{\left| \omega_{\lambda} \right|}{\omega_m} \right]^{\frac13}. \lb{zm}
\ee
It is seen from \eqref{zl} that $z_{\lambda}\to 0$ as $|\omega_{\lambda}|$ increases and $w_X$ is more phantom. When this is the case there might be some parameter values for which $H^2(z)$ itself becomes negative for some redshifts in the range $z_{\lambda}<z<z_m$, such models are not viable and must be rejected. The condition $H^2(z)>0$ is easily translated into 
\be
\omega_m \left[ (1+z)^3 - f_X(z) \right] + h^2 f_X(z) > \left| \omega_{\lambda} \right| 
                \left( 1 - f_X(z) \right). \lb{H2pos} 
 \ee
This inequality depends on the free parameters $w_X,~\omega_{\lambda},~h$. However, we should remember that in our theoretical analysis once $w_X$ and $\omega_{\lambda}$ are given, because of the constraint \eqref{constraint} $h$ is no longer free as we illustrate with Figs.~\ref{Fig3} and \ref{Fig4}. We see that \emph{a priori}, for given $w_X$, large$\left|\omega_{\lambda}\right|$ and low $h^2$ values can lead to a violation of \eqref{H2pos}.
Once a two-dimensional surface $h(w_X,~\omega_{\lambda})$ is found satisfying \eqref{H2pos} (see the lower panel of Fig.~\ref{Fig3}), the projection in the $(w_X,~\omega_{\lambda})$ plane satisfies it automatically too.

\section{Comparison with data: Model Selection and Parameter Estimation}
In this section, we are interested in the question of model selection: namely, comparing the different models to the reference \lcdm model. 
Let $M$ be the model, $D$ the data, and $\Theta$ the parameters of the model. 
Bayes theorem can be written as 
\begin{align}
    \Pr(\Theta|M,D) & = \frac{\Pr(D|\Theta,M)\Pr(\Theta|M)}{\Pr(D|M)},
\end{align}
where $\mathcal{P}(\Theta) = \Pr(\Theta|M,D)$ is the posterior distribution, $\mathcal{L}(\Theta)= \Pr(D|\Theta,M) $ is the likelihood, 
$\pi(\Theta) = \Pr(\Theta|M)$ is the prior, and  $\mathcal{Z} = \Pr(D|M)$ is the evidence.

\begin{table}[t]
    \centering
    \begin{tabular}{ccc}
    \toprule
Models &     Parameter & Prior range\\
    \midrule
   All &      $h$       & $[0.5,1]$\\
       $\lambda$\,all & $\omega_\lambda$ &  $[-4,0]$ \\
($\lambda$)$w$/I/III/CPL$w_0$    &     $w_0$    &  $[-1.2,-0.8]$\\
($\lambda$)II/III    &     $w_\infty$     &  $[-1.2,-0.8]$\\
        ($\lambda$)CPL$w_a$ &  $w_a$ & $[-0.2,0.2]$\\
    \bottomrule
    \end{tabular} 
    \caption{Flat prior range used in the nested sampling.  
    For each model, ($\lambda$) denotes both cases with and without $\lambda$}
    \label{tab:prior}
\end{table}

For parameter evaluation within a given model, the evidence can be seen as a normalization constant, and thus ignored. 
However, in order to perform model selection, the Bayesian evidences of the models have to be evaluated and compared. 
Calculating the evidence can be computationally challenging, in particular when using Monte Carlo Markov Chains. 
Therefore, in order to calculate the posterior distributions and the evidence, we use the nested-sampling algorithm \cite{2004AIPC..735..395S} as implemented in \texttt{pymultinest} \cite{2009MNRAS.398.1601F, 2014A&A...564A.125B}. 
We follow the authors' recommendations and use different sampling efficiencies for evidence evaluation and parameter estimation (0.3 and 0.8, respectively). 
We used 1000 live points and a tolerance factor of 1. 
We checked that the tolerance factor does not affect too much the results.

\begin{figure*}[t]
    {\includegraphics[width=0.495\textwidth]{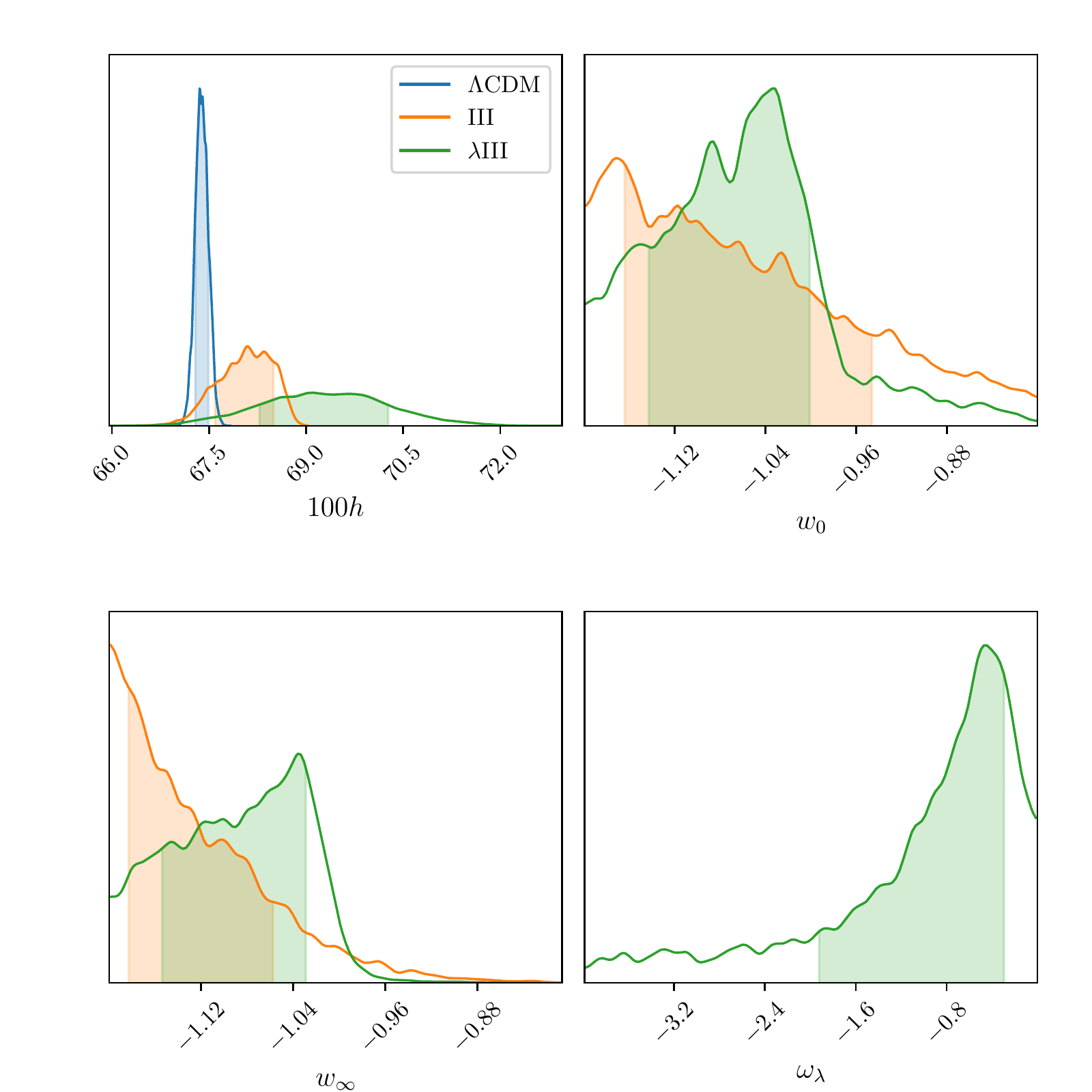}}~
    \hspace{0cm}
    {\includegraphics[width=0.495\textwidth]{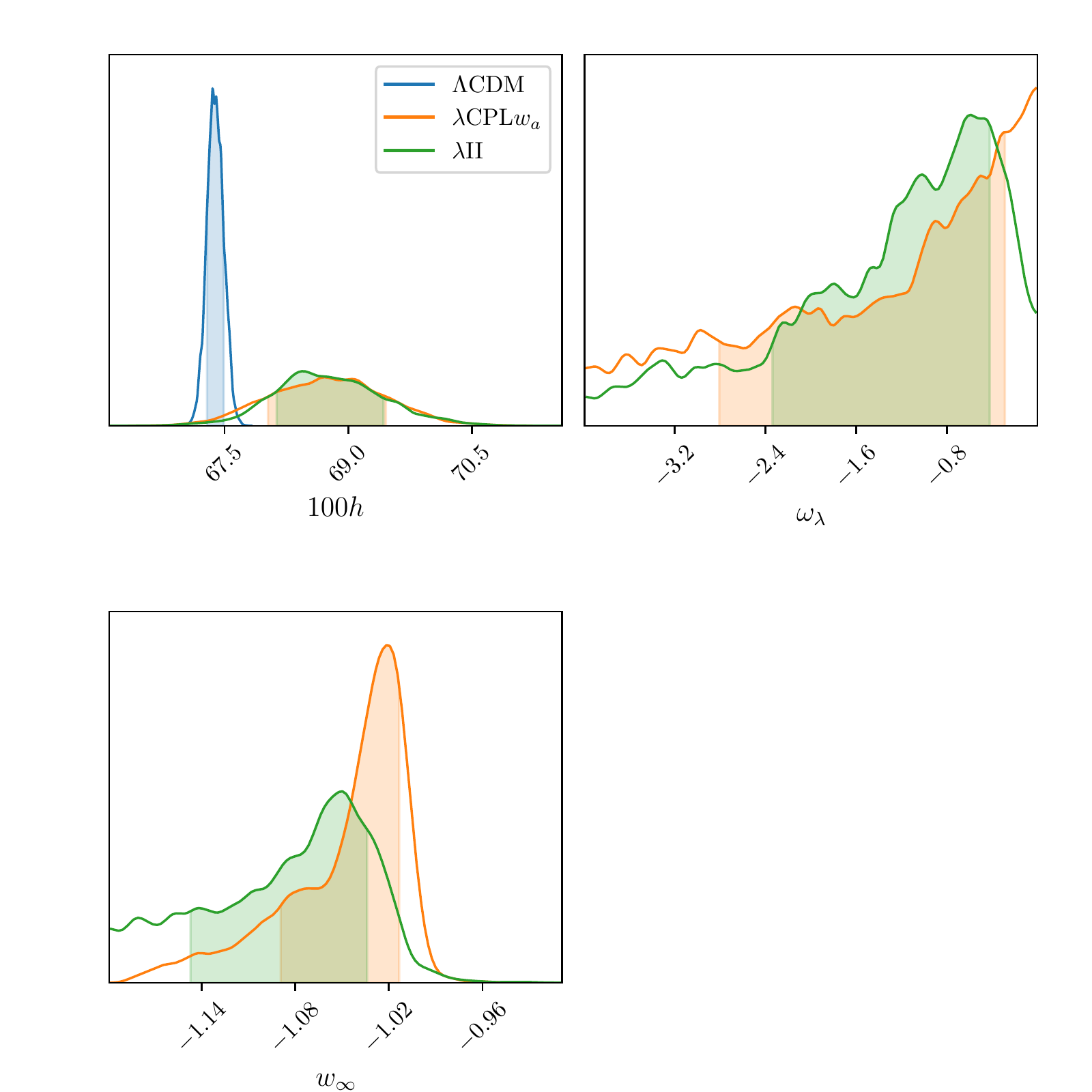}}
    \caption{Marginalized posteriors for different models using SNIa, BAO, $\ts$, and $H_0$ from HST Mira. The central value is the median, and the shared areas show the 68\% credible intervals around it. Left: \lcdm (blue), III (orange), and $\lambda$III (green). Right: \lcdm (blue), $\lambda$CPL$w_a$ (orange), $\lambda$II (green).}
\label{fig:postIII}
\end{figure*}

The priors are shown in Table~\ref{tab:prior}.  
As for $h$, because of its crucial role in this work, we choose 
rather wide, uninformative priors. 
For the other parameters associated to dark energy, we base our priors on physical properties of existing dark energy models, for example the difficulty to obtain realistic phantom models with $w\ll -1$.
Finally, we take the priors $\omega_\lambda \in [-4,0]$; 
see Figs.~\ref{Fig3} and \ref{Fig4} of Sec.~\ref{sec:h0}.

An elegant way to compare two models (say model $M_1$ and the null hypothesis $M_0$)  with different degrees of freedom is to compute the posterior ratio
\begin{align}
    \frac{\Pr(M_1|D)}{\Pr(M_0|D)} & = \frac{\Pr(D|M_1)\Pr(M_1)}{\Pr(D|M_0)\Pr(M_0)}\\
    & = \frac{\mathcal{Z}_1}{\mathcal{Z}_0} \equiv K\label{eq:K}.
\end{align}
In \eqref{eq:K}, we assume $\Pr(M_1) = \Pr(M_0)$.
Therefore, $K>1$ shows a preference for model 1 over model 0. 
The value of $K$ gives the degree of preference for one model over the other. 
Several  scales have been used, such as the Jeffreys scale, although the latter should be taken as an indication and interpreted with caution.

\paragraph{The data}

{The quantities  $\omega_\mathrm{m}$ and $\omega_\mathrm{r}$ are kept to their fiducial Planck value, and thus so are $r_\text{s}$ and $r_\text{d}$.
We vary $h$ and the parameters $\omega_\lambda,~w_X$ associated with dark energy and obtain $\theta_\mathrm{s,model}$ for each set of free parameters which is then compared to $\theta_\mathrm{s,Planck}$ and gives us a likelihood for $\theta_\text{s}$.  
}
In addition, we used  the 1048 distance modulus measurements from the Pantheon type Ia supernovae (SNIa) compilation \cite{2018ApJ...859..101S}, and the BAO from the Baryon Oscillation Spectroscopic Survey (BOSS) and extended-BOSS (eBOSS) surveys \cite{2017MNRAS.470.2617A,2017MNRAS.466..762Z}. 
Since the measurements from the SH0ES project are in tension with \lcdm, we instead use a measurement with larger error bars which is not inconsistent \cite{2020ApJ...889....5H}. 
In general, supernovae show a degeneracy between the absolute magnitude and the Hubble constant. 
Therefore, SNIa alone cannot measure $H_0$. 
However, in this particular study, since we fixed $\omega_m$, choosing a certain value for $\Omega_\mathrm{m,0}$ uniquely fixes $h$; therefore we can use SNIa data to obtain $h$.
The BAO provide us with $H(z)\rd$ and $d_\mathrm{A}(z)/\rd$, where $d_\mathrm{A}$ is the angular diameter distance and the sound horizon at the drag epoch  \rd is fixed to its Planck value (Table~\ref{tab:fix}). 
The three pairs of data points ($\dang/\rd$ and $H\rd$) from BOSS are correlated, and so are the four eBOSS pairs, and the BAO covariance matrix is thus
\begin{align}
    \mathbf{C}_\text{BAO} & =\begin{pmatrix}
\mathbf{C}_\text{BOSS} & 0\\
0 & \mathbf{C}_\text{eBOSS}
\end{pmatrix}.
\end{align}
$\ts$ and $H_0$ are one data point each, and their associated likelihood is thus trivial.
We used flat priors as shown in Table~\ref{tab:prior}.

\begin{table*}
    \centering
    \caption{68\% credible intervals for SN+BAO+\ts. }
    \label{tab:K}
    \begin{tabular}{ccccccc}
        \toprule
        Model & $100h$ & $w_0$ & $w_\infty$ or $w_a$  & $\omega_{\lambda}$ & $\ln \mathcal Z$ & $K$ \\ 
        \midrule
        $\Lambda$ & $67.382^{+0.107}_{-0.096}$ &  &   &  &  $-532.4$ & 1 \\ 
        \midrule
        $w$ & $68.62^{+0.85}_{-0.83}$ & $-1.042^{+0.028}_{-0.029}$ &  & $[0]$ & $-533.1$ &0.54  \\ 
        $\lambda w$ & $68.66^{+0.79}_{-0.75}$ & $-1.0102^{+0.0076}_{-0.0203}$ & & $-0.91^{+0.73}_{-1.88}$ & $-534.7$ & 0.11 \\ 
        \midrule
        CPL & $68.60^{+0.78}_{-0.84}$ & $-1.038^{+0.032}_{-0.029}$  & $-0.020^{+0.074}_{-0.055}$ & $[0]$ & $-532.6 $ & 0.83 \\ 
        $\lambda$CPL & $68.72^{+0.77}_{-0.87}$ & $-1.007^{+0.016}_{-0.028}$ & $-0.58^{+0.43}_{-1.11}$ & $-0.032^{+0.058}_{-0.044}$ & $-534.1 $ &  0.18\\ 
        \midrule
        CPL$w_0$ & $68.52^{+0.83}_{-0.85}$ & $-1.052^{+0.039}_{-0.038}$ & [$-(1+w_0)$] &$[0]$ & $-533.0 $ &  0.60\\ 
        $\lambda$CPL$w_0$ & $68.51^{+0.72}_{-0.74}$ & $-1.014^{+0.011}_{-0.021}$ & [$-(1+w_0)$] & $-0.84^{+0.65}_{-1.57}$ & $-534.6$ & 0.12 \\ 
        \midrule
        CPL$w_a$ & $68.34^{+0.39}_{-0.63}$ & $[-1]$ & $-0.124^{+0.083}_{-0.050}$ & $[0]$ &  $-532.1$ &1.48  \\ 
        $\lambda$CPL$w_a$ & $68.63^{+0.78}_{-0.75}$ &$[-1]$   &  $-0.034^{+0.024}_{-0.057}$ & $-1.03^{+0.78}_{-1.81}$ & $-533.1$ & 0.52\\ 
        \midrule
        I & $68.46^{+0.82}_{-0.83}$ & $-1.042^{+0.033}_{-0.032}$ & $[-1]$ & $[0]$ & $-533.2$ &0.48 \\ 
        $\lambda$I & $68.41^{+0.82}_{-0.75}$ & $-1.0092^{+0.0081}_{-0.0242}$ & $[-1]$ & $-0.92^{+0.76}_{-1.92}$ & $-534.8$ & 0.093 \\ 
        \midrule
        II & $67.86^{+0.21}_{-0.34}$ & $[-1]$ & $-1.127^{+0.090}_{-0.053}$ & $[0]$ &  $-532.2$ & 1.33 \\ 
        $\lambda$II & $68.58^{+0.66}_{-0.59}$ & $[-1]$ & $-1.059^{+0.035}_{-0.070}$ &  $-1.15^{+0.75}_{-1.76}$ &  $-532.1$ & 1.36\\ 
        \midrule
        III & $67.93^{+0.47}_{-0.50}$ &$-1.04^{+0.14}_{-0.11}$ &  $-1.128^{+0.092}_{-0.049}$ & $[0]$  &$-532.3$ & 1.20 \\ 
        $\lambda$III & $68.93^{+1.03}_{-0.89}$ & $-1.046^{+0.070}_{-0.083}$ & $-1.090^{+0.054}_{-0.068}$  & $-0.65^{+0.43}_{-0.95}$ &  $-531.4$  &2.96  \\ 
        \bottomrule
    \end{tabular}
\end{table*}

\begin{table*}
    \centering
    \caption{68\% credible intervals for SN+BAO+\ts+$H_0$ (HST Mira).}
    \label{tab:KH0}
    \begin{tabular}{ccccccc}
        \toprule
        Model & $100h$ & $w_0$ & $w_\infty$ or $w_a$  & $\omega_{\lambda}$ & $\ln Z$ & $K$\\ 
        \midrule
        $\Lambda$ & $67.386^{+0.105}_{-0.099}$ & & & &  $-533.5$ &1\\ 
        \midrule
        $w$ & $68.83^{+0.82}_{-0.83}$ &  $-1.049^{+0.028}_{-0.027}$ & & $[0]$ & $-533.7$  &0.82  \\ 
        $\lambda w$ & $68.81^{+0.86}_{-0.77}$  &  $-1.0122^{+0.0086}_{-0.0251}$ & & $-0.81^{+0.66}_{-1.75}$ &$-535.4$ &0.16 \\ 
        \midrule
        CPL & $68.77^{+0.75}_{-0.80}$ & $-1.044^{+0.031}_{-0.028}$ & $-0.020^{+0.072}_{-0.057}$ & $[0]$ &  $-533.3 $& 1.26\\ 
        $\lambda$CPL & $68.75^{+0.84}_{-0.78}$ & $-1.005^{+0.015}_{-0.022}$  & $-0.035^{+0.053}_{-0.042}$ &  $-0.71^{+0.52}_{-1.26}$ &  $-534.8$ & 0.30 \\ 
        \midrule
        CPL$w_0$ & $68.71^{+0.82}_{-0.83}$ & $-1.061^{+0.038}_{-0.037}$ & [$-(1+w_0)$]& $[0]$ & $-533.7$ &  0.87\\ 
        $\lambda$CPL$w_0$ & $68.72^{+0.83}_{-0.93}$  &  $-1.016^{+0.012}_{-0.033}$ & [$-(1+w_0)$] & $-0.72^{+0.58}_{-1.64}$ &$-535.3$  & 0.18\\ 
        \midrule
        CPL$w_a$ & $68.42^{+0.35}_{-0.48}$ & $[-1]$ &  $-0.134^{+0.063}_{-0.044}$ & $[0]$ & $-532.8$  & 1.99 \\ 
        $\lambda$CPL$w_a$ & $68.76^{+0.70}_{-0.73}$  & $[-1]$ &  $-0.035^{+0.022}_{-0.054}$ &  $-1.12^{+0.83}_{-1.69}$  &   $-533.7$ & 0.81  \\ 
        \midrule
        I & $68.67^{+0.80}_{-0.82}$ & $-1.050^{+0.032}_{-0.031}$ & $[-1]$ & $[0]$ &  $-533.9$   &  0.69  \\ 
        $\lambda$I & $68.51^{+1.09}_{-0.77}$  &$-1.0096^{+0.0076}_{-0.0272}$ & $[-1]$ & $-0.96^{+0.80}_{-1.92}$ & $-535.6$ & 0.13 \\ 
        \midrule
        II & $67.89^{+0.18}_{-0.30}$ & $[-1]$ &  $-1.137^{+0.083}_{-0.044}$ & $[0]$ & $-533.1$ &1.53 \\
        $\lambda$II & $68.73^{+0.70}_{-0.60}$ & $[-1]$ & $-1.074^{+0.040}_{-0.073}$ & $-1.09^{+0.67}_{-1.24}$ & $-532.6$ &2.54\\ 
        \midrule
        III &$68.09^{+0.41}_{-0.50}$  &  $-1.077^{+0.131}_{-0.087}$  &  $-1.137^{+0.080}_{-0.045}$  & $[0]$ &   $-533.1$ & 1.52  \\ 
        $\lambda$III & $69.28^{+0.99}_{-1.00}$  & $-1.063^{+0.062}_{-0.080}$ & $-1.083^{+0.054}_{-0.071}$ & $-0.74^{+0.45}_{-1.18}$ &  $-533.3$  &1.31 \\ 
        \bottomrule
    \end{tabular}
\end{table*}

Table~\ref{tab:K} summarizes our results  for SN+BAO+\ts\, while in Table~\ref{tab:KH0} the $H_0$ data point from HST-Mira is added.
The quoted central values and error bars correspond to the median and 68\% credible intervals around it. 
We remind the reader that in this analysis, $\omega_m$ , $\omega_b$, $N_\text{eff}$, and  $\sum m_\nu$ are fixed to their Planck value, and thus so are $\rs$ and $r_\mathrm{d}$ (see Table~\ref{tab:fix}). 

Figure~\ref{fig:postIII}
shows the posterior distributions of the parameters for models \lcdm, III, and $\lambda$III (left-hand panel),  
and \lcdm, $\lambda$CPL$w_a$, and $\lambda$II (right-hand panel). 
Model III shows preference for a higher value of $h$, and adding a negative $\lambda$ pushes $h$ even higher, although this is not sufficient to reconcile it with the SH0ES value.
An interesting property of the negative cosmological constant is that it allows to satisfy the observational constraints with an equation of state which is less phantom. 
This is clearly seen in particular for scenario $w$, where the addition of $\lambda$ shifts the central value of $w_0$ from $-1.049$ to $-1.0122$. 
We note that given our priors on $w_0$ and $w_\infty$, the equation of state is not always constrained, as seen, for instance, for model III. 
It is also interesting that the best models $\lambda$II are those where the phantom behavior, and hence the departure from $\Lambda$CDM, takes place at $z\gtrsim 1$. 
This suggests new physics appearing at these redshifts rather than at redshifts $z\lesssim 1$ can yield viable models. 
Models ($\lambda$)III fare reasonably well while having an additional departure from $\Lambda$CDM at very low redshifts $z\leq 0.1$ but it is clear when we compare them with models ($\lambda$)I that their main advantage comes from their phantomness at $z\gtrsim 1$. 
This seems further supported by the better evidence for ($\lambda$)CPL$w_a$, where $w_0=-1$ with a phantom behavior at higher redshifts ($w_a<0$), compared to ($\lambda$)CPL$w_0$ with $w$ tending asymptotically to $-1$ and departure from $-1$ takes place essentially at low redshifts. 
Note that CPL$w_a$ models have the best evidence for $\lambda=0$,  while models II have the highest evidence for $\lambda\ne 0$ ($\lambda$II). Interestingly, $\lambda$CPL$w_a$ lowers the evidence compared to 
CPL$w_a$. As we have said earlier, perhaps with the exception of the models $\lambda$II, the higher evidence 
compared to $\Lambda$CDM is not decisive and should be interpreted with caution.

\section{Summary and Conclusion}
In this work, we have considered the possibility that the dark energy sector contains a negative cosmological constant $\lambda$. Indeed, theoretical considerations from high energy physics suggest the possible presence of a negative cosmological constant rather than a positive one. This constitutes a radical change as in that case, and contrary to a positive cosmological constant $\Lambda$, this constant cannot produce the late-time accelerated expansion rate and a more sophisticated dark sector is required. The Universe acceleration is produced here by the X component. Clearly the presence of $\lambda$ can affect the expansion history and we have studied the viability of these models, also when a high $H_0$ is considered. 

While as we have shown some models can achieve a higher $H_0$ when the equation 
of state of the X component $w_X$ is of the phantom type--this is of course a generic property not restricted to $\lambda\ne 0$--we have investigated whether these models are viable when SNIa and BAO data are taken into account. We find indeed that most of our models are viable with a fair evidence for the models $\lambda$II.
Taking into account the $H_0$ value of the HST-Mira experiment reinforces the evidence of the models $\lambda$II, 
reaching a value $h\approx 0.7$ but not higher. Hence, while these models do alleviate the $H_0$ tension, a value for $H_0$ substantially higher would rule them out. We note also that the CPL$w_a$ models are the best models for $\lambda=0$ while the presence of a nonvanishing $\lambda$ lowers the evidence for $\lambda$CPL$w_a$ versus CPL$w_a$. 
It is further interesting that the best models $\lambda$II are equal to $\Lambda$CDM on $z\le 1$ and of the phantom type only at higher redshifts. 

The constant $\lambda$ will manifest itself in a very explicit way in another context, namely, the future evolution of our Universe. Considering for concreteness a constant $w_X$, it is clear that an equation of state $-1 < w_X$, sufficiently negative in order to produce an accelerated expansion 
rate today, will necessarily produce a transient acceleration stage. It will then eventually lead to a recollapsing  universe. 
We have found the analytical expression for the scale factor $a(t)$ in the regime around the turning 
point when dustlike matter is negligible compared to the dark sector. On the other hand if the X component is of the phantom type, $w_X<-1$, our Universe will end in a Big Rip as expected and we have found here too an analytical fit for $a(t)$ valid in the asymptotic region when dustlike matter becomes negligible compared to $\lambda$ while the latter is not yet negligible compared to the X component. 

Suggested by high energy physics, the possibility to have a negative cosmological constant is worth investigating as it challenges our intuition about the phenomenology of cosmological models. 
If this negative cosmological constant is substantial enough to affect the cosmic expansion history like in those models investigated here, a high value for $H_0$ could be a decisive test.

\section*{Acknowledgments}
The work of R.G. is supported by FONDECYT Project No 1171384. B.~L. would like to acknowledge the support of the National Research Foundation of Korea (Grant No. NRF-2019R1I1A1A01063740).
This work benefited from a high performance computing cluster (Seondeok) at the Korea Astronomy and Space Science Institute.
This research made use of Astropy, a community-developed core Python package for Astronomy \citep{2018AJ....156..123A, 2013A&A...558A..33A}, ChainConsumer \cite{Hinton2016}, matplotlib, a Python library for publication quality graphics \citep{Hunter:2007}, SciPy \citep{Virtanen_2020}, and NumPy \citep{van2011numpy}. 


\bibliography{biblio}

\begin{thebibliography}{61}%
\makeatletter
\providecommand \@ifxundefined [1]{%
 \@ifx{#1\undefined}
}%
\providecommand \@ifnum [1]{%
 \ifnum #1\expandafter \@firstoftwo
 \else \expandafter \@secondoftwo
 \fi
}%
\providecommand \@ifx [1]{%
 \ifx #1\expandafter \@firstoftwo
 \else \expandafter \@secondoftwo
 \fi
}%
\providecommand \natexlab [1]{#1}%
\providecommand \enquote  [1]{``#1''}%
\providecommand \bibnamefont  [1]{#1}%
\providecommand \bibfnamefont [1]{#1}%
\providecommand \citenamefont [1]{#1}%
\providecommand \href@noop [0]{\@secondoftwo}%
\providecommand \href [0]{\begingroup \@sanitize@url \@href}%
\providecommand \@href[1]{\@@startlink{#1}\@@href}%
\providecommand \@@href[1]{\endgroup#1\@@endlink}%
\providecommand \@sanitize@url [0]{\catcode `\\12\catcode `\$12\catcode
  `\&12\catcode `\#12\catcode `\^12\catcode `\_12\catcode `\%12\relax}%
\providecommand \@@startlink[1]{}%
\providecommand \@@endlink[0]{}%
\providecommand \url  [0]{\begingroup\@sanitize@url \@url }%
\providecommand \@url [1]{\endgroup\@href {#1}{\urlprefix }}%
\providecommand \urlprefix  [0]{URL }%
\providecommand \Eprint [0]{\href }%
\providecommand \doibase [0]{https://doi.org/}%
\providecommand \selectlanguage [0]{\@gobble}%
\providecommand \bibinfo  [0]{\@secondoftwo}%
\providecommand \bibfield  [0]{\@secondoftwo}%
\providecommand \translation [1]{[#1]}%
\providecommand \BibitemOpen [0]{}%
\providecommand \bibitemStop [0]{}%
\providecommand \bibitemNoStop [0]{.\EOS\space}%
\providecommand \EOS [0]{\spacefactor3000\relax}%
\providecommand \BibitemShut  [1]{\csname bibitem#1\endcsname}%
\let\auto@bib@innerbib\@empty
\bibitem [{\citenamefont {{Sahni}}\ and\ \citenamefont
  {{Starobinsky}}(2000)}]{2000IJMPD...9..373S}%
  \BibitemOpen
  \bibfield  {author} {\bibinfo {author} {\bibfnamefont {V.}~\bibnamefont
  {{Sahni}}}\ and\ \bibinfo {author} {\bibfnamefont {A.}~\bibnamefont
  {{Starobinsky}}},\ }\bibfield  {title} {\bibinfo {title} {{The Case for a
  Positive Cosmological {\ensuremath{\Lambda}}-Term}},\ }\href
  {https://doi.org/10.1142/S0218271800000542} {\bibfield  {journal} {\bibinfo
  {journal} {Int. J. Mod. Phys. D}\ }\textbf {\bibinfo {volume} {09}},\
  \bibinfo {pages} {373} (\bibinfo {year} {2000})}\BibitemShut {NoStop}%
\bibitem [{\citenamefont {Peebles}\ and\ \citenamefont
  {Ratra}(2003)}]{Peebles:2002gy}%
  \BibitemOpen
  \bibfield  {author} {\bibinfo {author} {\bibfnamefont {P.~J.~E.}\
  \bibnamefont {Peebles}}\ and\ \bibinfo {author} {\bibfnamefont
  {B.}~\bibnamefont {Ratra}},\ }\bibfield  {title} {\bibinfo {title} {{The
  Cosmological Constant and Dark Energy}},\ }\href
  {https://doi.org/10.1103/RevModPhys.75.559} {\bibfield  {journal} {\bibinfo
  {journal} {Rev. Mod. Phys.}\ }\textbf {\bibinfo {volume} {75}},\ \bibinfo
  {pages} {559} (\bibinfo {year} {2003})}\BibitemShut {NoStop}%
\bibitem [{\citenamefont {Padmanabhan}(2003)}]{Padmanabhan:2002ji}%
  \BibitemOpen
  \bibfield  {author} {\bibinfo {author} {\bibfnamefont {T.}~\bibnamefont
  {Padmanabhan}},\ }\bibfield  {title} {\bibinfo {title} {{Cosmological
  constant: The Weight of the vacuum}},\ }\href
  {https://doi.org/10.1016/S0370-1573(03)00120-0} {\bibfield  {journal}
  {\bibinfo  {journal} {\physrep}\ }\textbf {\bibinfo {volume} {380}},\
  \bibinfo {pages} {235} (\bibinfo {year} {2003})}\BibitemShut {NoStop}%
\bibitem [{\citenamefont {Copeland}\ \emph {et~al.}(2006)\citenamefont
  {Copeland}, \citenamefont {Sami},\ and\ \citenamefont
  {Tsujikawa}}]{Copeland:2006wr}%
  \BibitemOpen
  \bibfield  {author} {\bibinfo {author} {\bibfnamefont {E.~J.}\ \bibnamefont
  {Copeland}}, \bibinfo {author} {\bibfnamefont {M.}~\bibnamefont {Sami}},\
  and\ \bibinfo {author} {\bibfnamefont {S.}~\bibnamefont {Tsujikawa}},\
  }\bibfield  {title} {\bibinfo {title} {{Dynamics of dark energy}},\ }\href
  {https://doi.org/10.1142/S021827180600942X} {\bibfield  {journal} {\bibinfo
  {journal} {Int. J. Mod. Phys.}\ }\textbf {\bibinfo {volume} {D15}},\ \bibinfo
  {pages} {1753} (\bibinfo {year} {2006})}\BibitemShut {NoStop}%
\bibitem [{\citenamefont {Weinberg}\ \emph {et~al.}(2013)\citenamefont
  {Weinberg}, \citenamefont {Mortonson}, \citenamefont {Eisenstein},
  \citenamefont {Hirata}, \citenamefont {Riess},\ and\ \citenamefont
  {Rozo}}]{Weinberg:2012es}%
  \BibitemOpen
  \bibfield  {author} {\bibinfo {author} {\bibfnamefont {D.~H.}\ \bibnamefont
  {Weinberg}}, \bibinfo {author} {\bibfnamefont {M.~J.}\ \bibnamefont
  {Mortonson}}, \bibinfo {author} {\bibfnamefont {D.~J.}\ \bibnamefont
  {Eisenstein}}, \bibinfo {author} {\bibfnamefont {C.}~\bibnamefont {Hirata}},
  \bibinfo {author} {\bibfnamefont {A.~G.}\ \bibnamefont {Riess}},\ and\
  \bibinfo {author} {\bibfnamefont {E.}~\bibnamefont {Rozo}},\ }\bibfield
  {title} {\bibinfo {title} {{Observational Probes of Cosmic Acceleration}},\
  }\href {https://doi.org/10.1016/j.physrep.2013.05.001} {\bibfield  {journal}
  {\bibinfo  {journal} {\physrep}\ }\textbf {\bibinfo {volume} {530}},\
  \bibinfo {pages} {87} (\bibinfo {year} {2013})}\BibitemShut {NoStop}%
\bibitem [{\citenamefont {Amendola}\ \emph {et~al.}(2013)\citenamefont
  {Amendola} \emph {et~al.}}]{Amendola:2012ys}%
  \BibitemOpen
  \bibfield  {author} {\bibinfo {author} {\bibfnamefont {L.}~\bibnamefont
  {Amendola}} \emph {et~al.} (\bibinfo {collaboration} {Euclid Theory Working
  Group}),\ }\bibfield  {title} {\bibinfo {title} {{Cosmology and fundamental
  physics with the Euclid satellite}},\ }\href
  {https://doi.org/10.12942/lrr-2013-6} {\bibfield  {journal} {\bibinfo
  {journal} {Living Rev. Relativity}\ }\textbf {\bibinfo {volume} {16}},\
  \bibinfo {pages} {6} (\bibinfo {year} {2013})}\BibitemShut {NoStop}%
\bibitem [{\citenamefont {Bullock}\ and\ \citenamefont
  {Boylan-Kolchin}(2017)}]{Bullock:2017xww}%
  \BibitemOpen
  \bibfield  {author} {\bibinfo {author} {\bibfnamefont {J.~S.}\ \bibnamefont
  {Bullock}}\ and\ \bibinfo {author} {\bibfnamefont {M.}~\bibnamefont
  {Boylan-Kolchin}},\ }\bibfield  {title} {\bibinfo {title} {{Small-Scale
  Challenges to the $\Lambda$CDM Paradigm}},\ }\href
  {https://doi.org/10.1146/annurev-astro-091916-055313} {\bibfield  {journal}
  {\bibinfo  {journal} {Annu. Rev. Astron. Astrophys.}\ }\textbf {\bibinfo
  {volume} {55}},\ \bibinfo {pages} {343} (\bibinfo {year} {2017})}\BibitemShut
  {NoStop}%
\bibitem [{\citenamefont {{Raveri}}\ and\ \citenamefont
  {{Hu}}(2019)}]{2019PhRvD..99d3506R}%
  \BibitemOpen
  \bibfield  {author} {\bibinfo {author} {\bibfnamefont {M.}~\bibnamefont
  {{Raveri}}}\ and\ \bibinfo {author} {\bibfnamefont {W.}~\bibnamefont
  {{Hu}}},\ }\bibfield  {title} {\bibinfo {title} {{Concordance and discordance
  in cosmology}},\ }\href {https://doi.org/10.1103/PhysRevD.99.043506}
  {\bibfield  {journal} {\bibinfo  {journal} {\prd}\ }\textbf {\bibinfo
  {volume} {99}},\ \bibinfo {eid} {043506} (\bibinfo {year}
  {2019})}\BibitemShut {NoStop}%
\bibitem [{\citenamefont {Verde}\ \emph {et~al.}(2017)\citenamefont {Verde},
  \citenamefont {Bernal}, \citenamefont {Heavens},\ and\ \citenamefont
  {Jimenez}}]{Verde:2016ccp}%
  \BibitemOpen
  \bibfield  {author} {\bibinfo {author} {\bibfnamefont {L.}~\bibnamefont
  {Verde}}, \bibinfo {author} {\bibfnamefont {J.~L.}\ \bibnamefont {Bernal}},
  \bibinfo {author} {\bibfnamefont {A.~F.}\ \bibnamefont {Heavens}},\ and\
  \bibinfo {author} {\bibfnamefont {R.}~\bibnamefont {Jimenez}},\ }\bibfield
  {title} {\bibinfo {title} {{The length of the low-redshift standard ruler}},\
  }\href {https://doi.org/10.1093/mnras/stx116} {\bibfield  {journal} {\bibinfo
   {journal} {\mnras}\ }\textbf {\bibinfo {volume} {467}},\ \bibinfo {pages}
  {731} (\bibinfo {year} {2017})}\BibitemShut {NoStop}%
\bibitem [{\citenamefont {Bernal}\ \emph {et~al.}(2016)\citenamefont {Bernal},
  \citenamefont {Verde},\ and\ \citenamefont {Riess}}]{Bernal:2016gxb}%
  \BibitemOpen
  \bibfield  {author} {\bibinfo {author} {\bibfnamefont {J.~L.}\ \bibnamefont
  {Bernal}}, \bibinfo {author} {\bibfnamefont {L.}~\bibnamefont {Verde}},\ and\
  \bibinfo {author} {\bibfnamefont {A.~G.}\ \bibnamefont {Riess}},\ }\bibfield
  {title} {\bibinfo {title} {{The trouble with $H_0$}},\ }\href
  {https://doi.org/10.1088/1475-7516/2016/10/019} {\bibfield  {journal}
  {\bibinfo  {journal} {\jcap}\ }\textbf {\bibinfo {volume} {10}},\ \bibinfo
  {pages} {019} (\bibinfo {year} {2016})}\BibitemShut {NoStop}%
\bibitem [{\citenamefont {{L'Huillier}}\ and\ \citenamefont
  {{Shafieloo}}(2017)}]{2017JCAP...01..015L}%
  \BibitemOpen
  \bibfield  {author} {\bibinfo {author} {\bibfnamefont {B.}~\bibnamefont
  {{L'Huillier}}}\ and\ \bibinfo {author} {\bibfnamefont {A.}~\bibnamefont
  {{Shafieloo}}},\ }\bibfield  {title} {\bibinfo {title} {{Model-independent
  test of the FLRW metric, the flatness of the Universe, and non-local
  estimation of H$_{0}$ r$_{d}$}},\ }\href
  {https://doi.org/10.1088/1475-7516/2017/01/015} {\bibfield  {journal}
  {\bibinfo  {journal} {\jcap}\ }\textbf {\bibinfo {volume} {2017}},\ \bibinfo
  {eid} {015} (\bibinfo {year} {2017})}\BibitemShut {NoStop}%
\bibitem [{\citenamefont {{Shafieloo}}\ \emph {et~al.}(2018)\citenamefont
  {{Shafieloo}}, \citenamefont {{L'Huillier}},\ and\ \citenamefont
  {{Starobinsky}}}]{2018PhRvD..98h3526S}%
  \BibitemOpen
  \bibfield  {author} {\bibinfo {author} {\bibfnamefont {A.}~\bibnamefont
  {{Shafieloo}}}, \bibinfo {author} {\bibfnamefont {B.}~\bibnamefont
  {{L'Huillier}}},\ and\ \bibinfo {author} {\bibfnamefont {A.~A.}\ \bibnamefont
  {{Starobinsky}}},\ }\bibfield  {title} {\bibinfo {title} {{Falsifying
  {\ensuremath{\Lambda}} CDM : Model-independent tests of the concordance model
  with eBOSS DR14Q and Pantheon}},\ }\href
  {https://doi.org/10.1103/PhysRevD.98.083526} {\bibfield  {journal} {\bibinfo
  {journal} {\prd}\ }\textbf {\bibinfo {volume} {98}},\ \bibinfo {eid} {083526}
  (\bibinfo {year} {2018})}\BibitemShut {NoStop}%
\bibitem [{\citenamefont {{Verde}}\ \emph {et~al.}(2019)\citenamefont
  {{Verde}}, \citenamefont {{Treu}},\ and\ \citenamefont
  {{Riess}}}]{2019NatAs...3..891V}%
  \BibitemOpen
  \bibfield  {author} {\bibinfo {author} {\bibfnamefont {L.}~\bibnamefont
  {{Verde}}}, \bibinfo {author} {\bibfnamefont {T.}~\bibnamefont {{Treu}}},\
  and\ \bibinfo {author} {\bibfnamefont {A.~G.}\ \bibnamefont {{Riess}}},\
  }\bibfield  {title} {\bibinfo {title} {{Tensions between the early and late
  Universe}},\ }\href {https://doi.org/10.1038/s41550-019-0902-0} {\bibfield
  {journal} {\bibinfo  {journal} {Nat. Astron.}\ }\textbf {\bibinfo {volume}
  {3}},\ \bibinfo {pages} {891} (\bibinfo {year} {2019})}\BibitemShut {NoStop}%
\bibitem [{\citenamefont {{Ade}}\ \emph {et~al.}(2016)\citenamefont {{Ade}},
  \citenamefont {{Aghanim}}, \citenamefont {{Arnaud}}, \citenamefont
  {{Ashdown}}, \citenamefont {{Aumont}}, \citenamefont {{Baccigalupi}},
  \citenamefont {{Banday}}, \citenamefont {{Barreiro}}, \citenamefont
  {{Bartlett}} \emph {et~al.}}]{2016A&A...594A..13P}%
  \BibitemOpen
  \bibfield  {author} {\bibinfo {author} {\bibfnamefont {P.~A.~R.}\
  \bibnamefont {{Ade}}}, \bibinfo {author} {\bibfnamefont {N.}~\bibnamefont
  {{Aghanim}}}, \bibinfo {author} {\bibfnamefont {M.}~\bibnamefont {{Arnaud}}},
  \bibinfo {author} {\bibfnamefont {M.}~\bibnamefont {{Ashdown}}}, \bibinfo
  {author} {\bibfnamefont {J.}~\bibnamefont {{Aumont}}}, \bibinfo {author}
  {\bibfnamefont {C.}~\bibnamefont {{Baccigalupi}}}, \bibinfo {author}
  {\bibfnamefont {A.~J.}\ \bibnamefont {{Banday}}}, \bibinfo {author}
  {\bibfnamefont {R.~B.}\ \bibnamefont {{Barreiro}}}, \bibinfo {author}
  {\bibfnamefont {J.~G.}\ \bibnamefont {{Bartlett}}}, \emph {et~al.} (\bibinfo
  {collaboration} {Planck Collaboration}),\ }\bibfield  {title} {\bibinfo
  {title} {{Planck 2015 results. XIII. Cosmological parameters}},\ }\href
  {https://doi.org/10.1051/0004-6361/201525830} {\bibfield  {journal} {\bibinfo
   {journal} {\aap}\ }\textbf {\bibinfo {volume} {594}},\ \bibinfo {eid} {A13}
  (\bibinfo {year} {2016})}\BibitemShut {NoStop}%
\bibitem [{\citenamefont {Cardenas}\ \emph {et~al.}(2003)\citenamefont
  {Cardenas}, \citenamefont {Gonzalez}, \citenamefont {Leiva}, \citenamefont
  {Martin},\ and\ \citenamefont {Quiros}}]{Cardenas:2002np}%
  \BibitemOpen
  \bibfield  {author} {\bibinfo {author} {\bibfnamefont {R.}~\bibnamefont
  {Cardenas}}, \bibinfo {author} {\bibfnamefont {T.}~\bibnamefont {Gonzalez}},
  \bibinfo {author} {\bibfnamefont {Y.}~\bibnamefont {Leiva}}, \bibinfo
  {author} {\bibfnamefont {O.}~\bibnamefont {Martin}},\ and\ \bibinfo {author}
  {\bibfnamefont {I.}~\bibnamefont {Quiros}},\ }\bibfield  {title} {\bibinfo
  {title} {{A model of the universe including dark energy accounted for by both
  a quintessence field and a (negative) cosmological constant}},\ }\href
  {https://doi.org/10.1103/PhysRevD.67.083501} {\bibfield  {journal} {\bibinfo
  {journal} {Phys. Rev. D}\ }\textbf {\bibinfo {volume} {67}},\ \bibinfo
  {pages} {083501} (\bibinfo {year} {2003})}\BibitemShut {NoStop}%
\bibitem [{\citenamefont {Visinelli}\ \emph {et~al.}(2019)\citenamefont
  {Visinelli}, \citenamefont {Vagnozzi},\ and\ \citenamefont
  {Danielsson}}]{Visinelli:2019qqu}%
  \BibitemOpen
  \bibfield  {author} {\bibinfo {author} {\bibfnamefont {L.}~\bibnamefont
  {Visinelli}}, \bibinfo {author} {\bibfnamefont {S.}~\bibnamefont
  {Vagnozzi}},\ and\ \bibinfo {author} {\bibfnamefont {U.}~\bibnamefont
  {Danielsson}},\ }\bibfield  {title} {\bibinfo {title} {{Revisiting a negative
  cosmological constant from low-redshift data}},\ }\href
  {https://doi.org/10.3390/sym11081035} {\bibfield  {journal} {\bibinfo
  {journal} {Symmetry}\ }\textbf {\bibinfo {volume} {11}},\ \bibinfo {pages}
  {1035} (\bibinfo {year} {2019})}\BibitemShut {NoStop}%
\bibitem [{\citenamefont {Ye}\ and\ \citenamefont {Piao}(2020)}]{Ye:2020btb}%
  \BibitemOpen
  \bibfield  {author} {\bibinfo {author} {\bibfnamefont {G.}~\bibnamefont
  {Ye}}\ and\ \bibinfo {author} {\bibfnamefont {Y.-S.}\ \bibnamefont {Piao}},\
  }\bibfield  {title} {\bibinfo {title} {{Is the Hubble tension a hint of AdS
  phase around recombination?}},\ }\href
  {https://doi.org/10.1103/PhysRevD.101.083507} {\bibfield  {journal} {\bibinfo
   {journal} {\prd}\ }\textbf {\bibinfo {volume} {101}},\ \bibinfo {pages}
  {083507} (\bibinfo {year} {2020})}\BibitemShut {NoStop}%
\bibitem [{\citenamefont {Akarsu}\ \emph {et~al.}(2020)\citenamefont {Akarsu},
  \citenamefont {Barrow}, \citenamefont {Escamilla},\ and\ \citenamefont
  {Vazquez}}]{Akarsu:2019hmw}%
  \BibitemOpen
  \bibfield  {author} {\bibinfo {author} {\bibfnamefont {{\"O}.}~\bibnamefont
  {Akarsu}}, \bibinfo {author} {\bibfnamefont {J.~D.}\ \bibnamefont {Barrow}},
  \bibinfo {author} {\bibfnamefont {L.~A.}\ \bibnamefont {Escamilla}},\ and\
  \bibinfo {author} {\bibfnamefont {J.~A.}\ \bibnamefont {Vazquez}},\
  }\bibfield  {title} {\bibinfo {title} {{Graduated dark energy: Observational
  hints of a spontaneous sign switch in the cosmological constant}},\ }\href
  {https://doi.org/10.1103/PhysRevD.101.063528} {\bibfield  {journal} {\bibinfo
   {journal} {\prd}\ }\textbf {\bibinfo {volume} {101}},\ \bibinfo {pages}
  {063528} (\bibinfo {year} {2020})}\BibitemShut {NoStop}%
\bibitem [{\citenamefont {Boisseau}\ \emph {et~al.}(2015)\citenamefont
  {Boisseau}, \citenamefont {Giacomini}, \citenamefont {Polarski},\ and\
  \citenamefont {Starobinsky}}]{Boisseau:2015hqa}%
  \BibitemOpen
  \bibfield  {author} {\bibinfo {author} {\bibfnamefont {B.}~\bibnamefont
  {Boisseau}}, \bibinfo {author} {\bibfnamefont {H.}~\bibnamefont {Giacomini}},
  \bibinfo {author} {\bibfnamefont {D.}~\bibnamefont {Polarski}},\ and\
  \bibinfo {author} {\bibfnamefont {A.}~\bibnamefont {Starobinsky}},\
  }\bibfield  {title} {\bibinfo {title} {{Bouncing Universes in Scalar-Tensor
  Gravity Models admitting Negative Potentials}},\ }\href
  {https://doi.org/10.1088/1475-7516/2015/07/002} {\bibfield  {journal}
  {\bibinfo  {journal} {\jcap}\ }\textbf {\bibinfo {volume} {07}},\ \bibinfo
  {pages} {002} (\bibinfo {year} {2015})}\BibitemShut {NoStop}%
\bibitem [{\citenamefont {Vazquez}\ \emph {et~al.}(2018)\citenamefont
  {Vazquez}, \citenamefont {Hee}, \citenamefont {Hobson}, \citenamefont
  {Lasenby}, \citenamefont {Ibison},\ and\ \citenamefont
  {Bridges}}]{Vazquez:2012ag}%
  \BibitemOpen
  \bibfield  {author} {\bibinfo {author} {\bibfnamefont {J.~A.}\ \bibnamefont
  {Vazquez}}, \bibinfo {author} {\bibfnamefont {S.}~\bibnamefont {Hee}},
  \bibinfo {author} {\bibfnamefont {M.~P.}\ \bibnamefont {Hobson}}, \bibinfo
  {author} {\bibfnamefont {A.~N.}\ \bibnamefont {Lasenby}}, \bibinfo {author}
  {\bibfnamefont {M.}~\bibnamefont {Ibison}},\ and\ \bibinfo {author}
  {\bibfnamefont {M.}~\bibnamefont {Bridges}},\ }\bibfield  {title} {\bibinfo
  {title} {{Observational constraints on conformal time symmetry, missing
  matter and double dark energy}},\ }\href
  {https://doi.org/10.1088/1475-7516/2018/07/062} {\bibfield  {journal}
  {\bibinfo  {journal} {\jcap}\ }\textbf {\bibinfo {volume} {07}},\ \bibinfo
  {pages} {062} (\bibinfo {year} {2018})}\BibitemShut {NoStop}%
\bibitem [{\citenamefont {Grande}\ \emph {et~al.}(2006)\citenamefont {Grande},
  \citenamefont {Sola},\ and\ \citenamefont {Stefancic}}]{Grande:2006nn}%
  \BibitemOpen
  \bibfield  {author} {\bibinfo {author} {\bibfnamefont {J.}~\bibnamefont
  {Grande}}, \bibinfo {author} {\bibfnamefont {J.}~\bibnamefont {Sola}},\ and\
  \bibinfo {author} {\bibfnamefont {H.}~\bibnamefont {Stefancic}},\ }\bibfield
  {title} {\bibinfo {title} {{LXCDM: A Cosmon model solution to the
  cosmological coincidence problem?}},\ }\href
  {https://doi.org/10.1088/1475-7516/2006/08/011} {\bibfield  {journal}
  {\bibinfo  {journal} {\jcap}\ }\textbf {\bibinfo {volume} {08}},\ \bibinfo
  {pages} {011} (\bibinfo {year} {2006})}\BibitemShut {NoStop}%
\bibitem [{\citenamefont {Dutta}\ \emph {et~al.}(2020)\citenamefont {Dutta},
  \citenamefont {Ruchika}, \citenamefont {Roy}, \citenamefont {Sen},\ and\
  \citenamefont {Sheikh-Jabbari}}]{Dutta:2018vmq}%
  \BibitemOpen
  \bibfield  {author} {\bibinfo {author} {\bibfnamefont {K.}~\bibnamefont
  {Dutta}}, \bibinfo {author} {\bibnamefont {Ruchika}}, \bibinfo {author}
  {\bibfnamefont {A.}~\bibnamefont {Roy}}, \bibinfo {author} {\bibfnamefont
  {A.~A.}\ \bibnamefont {Sen}},\ and\ \bibinfo {author} {\bibfnamefont {M.~M.}\
  \bibnamefont {Sheikh-Jabbari}},\ }\bibfield  {title} {\bibinfo {title}
  {{Beyond $\Lambda $CDM with low and high redshift data: implications for dark
  energy}},\ }\href {https://doi.org/10.1007/s10714-020-2665-4} {\bibfield
  {journal} {\bibinfo  {journal} {Gen. Relativ. Gravit.}\ }\textbf {\bibinfo
  {volume} {52}},\ \bibinfo {pages} {15} (\bibinfo {year} {2020})}\BibitemShut
  {NoStop}%
\bibitem [{\citenamefont {Alestas}\ \emph {et~al.}(2020)\citenamefont
  {Alestas}, \citenamefont {Kazantzidis},\ and\ \citenamefont
  {Perivolaropoulos}}]{Alestas:2020mvb}%
  \BibitemOpen
  \bibfield  {author} {\bibinfo {author} {\bibfnamefont {G.}~\bibnamefont
  {Alestas}}, \bibinfo {author} {\bibfnamefont {L.}~\bibnamefont
  {Kazantzidis}},\ and\ \bibinfo {author} {\bibfnamefont {L.}~\bibnamefont
  {Perivolaropoulos}},\ }\bibfield  {title} {\bibinfo {title} {{$H_0$ tension,
  phantom dark energy, and cosmological parameter degeneracies}},\ }\href
  {https://doi.org/10.1103/PhysRevD.101.123516} {\bibfield  {journal} {\bibinfo
   {journal} {Phys. Rev. D}\ }\textbf {\bibinfo {volume} {101}},\ \bibinfo
  {pages} {123516} (\bibinfo {year} {2020})}\BibitemShut {NoStop}%
\bibitem [{\citenamefont {Ó~Colgáin}\ \emph {et~al.}(2019)\citenamefont
  {Ó~Colgáin}, \citenamefont {van Putten},\ and\ \citenamefont
  {Yavartanoo}}]{Colgain:2018wgk}%
  \BibitemOpen
  \bibfield  {author} {\bibinfo {author} {\bibfnamefont {E.}~\bibnamefont
  {Ó~Colgáin}}, \bibinfo {author} {\bibfnamefont {M.~H.}\ \bibnamefont {van
  Putten}},\ and\ \bibinfo {author} {\bibfnamefont {H.}~\bibnamefont
  {Yavartanoo}},\ }\bibfield  {title} {\bibinfo {title} {{de Sitter Swampland,
  $H_0$ tension \& observation}},\ }\href
  {https://doi.org/10.1016/j.physletb.2019.04.032} {\bibfield  {journal}
  {\bibinfo  {journal} {Phys. Lett. B}\ }\textbf {\bibinfo {volume} {793}},\
  \bibinfo {pages} {126} (\bibinfo {year} {2019})}\BibitemShut {NoStop}%
\bibitem [{\citenamefont {Poulin}\ \emph {et~al.}(2019)\citenamefont {Poulin},
  \citenamefont {Smith}, \citenamefont {Karwal},\ and\ \citenamefont
  {Kamionkowski}}]{Poulin:2018cxd}%
  \BibitemOpen
  \bibfield  {author} {\bibinfo {author} {\bibfnamefont {V.}~\bibnamefont
  {Poulin}}, \bibinfo {author} {\bibfnamefont {T.~L.}\ \bibnamefont {Smith}},
  \bibinfo {author} {\bibfnamefont {T.}~\bibnamefont {Karwal}},\ and\ \bibinfo
  {author} {\bibfnamefont {M.}~\bibnamefont {Kamionkowski}},\ }\bibfield
  {title} {\bibinfo {title} {{Early Dark Energy Can Resolve The Hubble
  Tension}},\ }\href {https://doi.org/10.1103/PhysRevLett.122.221301}
  {\bibfield  {journal} {\bibinfo  {journal} {Phys. Rev. Lett.}\ }\textbf
  {\bibinfo {volume} {122}},\ \bibinfo {pages} {221301} (\bibinfo {year}
  {2019})}\BibitemShut {NoStop}%
\bibitem [{\citenamefont {Vattis}\ \emph {et~al.}(2019)\citenamefont {Vattis},
  \citenamefont {Koushiappas},\ and\ \citenamefont {Loeb}}]{Vattis:2019efj}%
  \BibitemOpen
  \bibfield  {author} {\bibinfo {author} {\bibfnamefont {K.}~\bibnamefont
  {Vattis}}, \bibinfo {author} {\bibfnamefont {S.~M.}\ \bibnamefont
  {Koushiappas}},\ and\ \bibinfo {author} {\bibfnamefont {A.}~\bibnamefont
  {Loeb}},\ }\bibfield  {title} {\bibinfo {title} {{Dark matter decaying in the
  late Universe can relieve the H0 tension}},\ }\href
  {https://doi.org/10.1103/PhysRevD.99.121302} {\bibfield  {journal} {\bibinfo
  {journal} {Phys. Rev. D}\ }\textbf {\bibinfo {volume} {99}},\ \bibinfo
  {pages} {121302} (\bibinfo {year} {2019})}\BibitemShut {NoStop}%
\bibitem [{\citenamefont {{Agrawal}}\ \emph {et~al.}()\citenamefont
  {{Agrawal}}, \citenamefont {{Obied}},\ and\ \citenamefont
  {{Vafa}}}]{Agrawal:2019dlm}%
  \BibitemOpen
  \bibfield  {author} {\bibinfo {author} {\bibfnamefont {P.}~\bibnamefont
  {{Agrawal}}}, \bibinfo {author} {\bibfnamefont {G.}~\bibnamefont {{Obied}}},\
  and\ \bibinfo {author} {\bibfnamefont {C.}~\bibnamefont {{Vafa}}},\
  }\href@noop {} {\bibinfo {title} {{$H_0$ Tension, Swampland Conjectures and
  the Epoch of Fading Dark Matter}}},\ \Eprint
  {https://arxiv.org/abs/1906.08261} {arXiv:1906.08261} \BibitemShut {NoStop}%
\bibitem [{\citenamefont {{Li}}\ and\ \citenamefont
  {{Shafieloo}}(2019)}]{2019ApJ...883L...3L}%
  \BibitemOpen
  \bibfield  {author} {\bibinfo {author} {\bibfnamefont {X.}~\bibnamefont
  {{Li}}}\ and\ \bibinfo {author} {\bibfnamefont {A.}~\bibnamefont
  {{Shafieloo}}},\ }\bibfield  {title} {\bibinfo {title} {{A Simple
  Phenomenological Emergent Dark Energy Model can Resolve the Hubble
  Tension}},\ }\href {https://doi.org/10.3847/2041-8213/ab3e09} {\bibfield
  {journal} {\bibinfo  {journal} {\apjl}\ }\textbf {\bibinfo {volume} {883}},\
  \bibinfo {eid} {L3} (\bibinfo {year} {2019})}\BibitemShut {NoStop}%
\bibitem [{\citenamefont {Vagnozzi}(2020)}]{Vagnozzi:2019ezj}%
  \BibitemOpen
  \bibfield  {author} {\bibinfo {author} {\bibfnamefont {S.}~\bibnamefont
  {Vagnozzi}},\ }\bibfield  {title} {\bibinfo {title} {{New physics in light of
  the $H_0$ tension: An alternative view}},\ }\href
  {https://doi.org/10.1103/PhysRevD.102.023518} {\bibfield  {journal} {\bibinfo
   {journal} {\prd}\ }\textbf {\bibinfo {volume} {102}},\ \bibinfo {pages}
  {023518} (\bibinfo {year} {2020})}\BibitemShut {NoStop}%
\bibitem [{\citenamefont {{Di Valentino}}\ \emph {et~al.}(2020)\citenamefont
  {{Di Valentino}}, \citenamefont {{Melchiorri}}, \citenamefont {{Mena}},\ and\
  \citenamefont {{Vagnozzi}}}]{2020PDU....3000666D}%
  \BibitemOpen
  \bibfield  {author} {\bibinfo {author} {\bibfnamefont {E.}~\bibnamefont {{Di
  Valentino}}}, \bibinfo {author} {\bibfnamefont {A.}~\bibnamefont
  {{Melchiorri}}}, \bibinfo {author} {\bibfnamefont {O.}~\bibnamefont
  {{Mena}}},\ and\ \bibinfo {author} {\bibfnamefont {S.}~\bibnamefont
  {{Vagnozzi}}},\ }\bibfield  {title} {\bibinfo {title} {{Interacting dark
  energy in the early 2020s: A promising solution to the H$_{0}$ and cosmic
  shear tensions}},\ }\href {https://doi.org/10.1016/j.dark.2020.100666}
  {\bibfield  {journal} {\bibinfo  {journal} {Phys. Dark Univ.}\ }\textbf
  {\bibinfo {volume} {30}},\ \bibinfo {eid} {100666} (\bibinfo {year}
  {2020})}\BibitemShut {NoStop}%
\bibitem [{\citenamefont {{Demianski}}\ \emph {et~al.}()\citenamefont
  {{Demianski}}, \citenamefont {{Piedipalumbo}}, \citenamefont {{Sawant}},\
  and\ \citenamefont {{Amati}}}]{Demianski:2019vzl}%
  \BibitemOpen
  \bibfield  {author} {\bibinfo {author} {\bibfnamefont {M.}~\bibnamefont
  {{Demianski}}}, \bibinfo {author} {\bibfnamefont {E.}~\bibnamefont
  {{Piedipalumbo}}}, \bibinfo {author} {\bibfnamefont {D.}~\bibnamefont
  {{Sawant}}},\ and\ \bibinfo {author} {\bibfnamefont {L.}~\bibnamefont
  {{Amati}}},\ }\href@noop {} {\bibinfo {title} {{High redshift constraints on
  dark energy models and tension with the flat LambdaCDM model}}},\ \Eprint
  {https://arxiv.org/abs/1911.08228} {arXiv:1911.08228} \BibitemShut {NoStop}%
\bibitem [{\citenamefont {Anchordoqui}\ \emph {et~al.}(2020)\citenamefont
  {Anchordoqui}, \citenamefont {Antoniadis}, \citenamefont {Lüst},
  \citenamefont {Soriano},\ and\ \citenamefont {Taylor}}]{Anchordoqui:2019amx}%
  \BibitemOpen
  \bibfield  {author} {\bibinfo {author} {\bibfnamefont {L.~A.}\ \bibnamefont
  {Anchordoqui}}, \bibinfo {author} {\bibfnamefont {I.}~\bibnamefont
  {Antoniadis}}, \bibinfo {author} {\bibfnamefont {D.}~\bibnamefont {Lüst}},
  \bibinfo {author} {\bibfnamefont {J.~F.}\ \bibnamefont {Soriano}},\ and\
  \bibinfo {author} {\bibfnamefont {T.~R.}\ \bibnamefont {Taylor}},\ }\bibfield
   {title} {\bibinfo {title} {{$H_0$ tension and the String Swampland}},\
  }\href {https://doi.org/10.1103/PhysRevD.101.083532} {\bibfield  {journal}
  {\bibinfo  {journal} {Phys. Rev. D}\ }\textbf {\bibinfo {volume} {101}},\
  \bibinfo {pages} {083532} (\bibinfo {year} {2020})}\BibitemShut {NoStop}%
\bibitem [{\citenamefont {{Benevento}}\ \emph {et~al.}(2020)\citenamefont
  {{Benevento}}, \citenamefont {{Hu}},\ and\ \citenamefont
  {{Raveri}}}]{2020PhRvD.101j3517B}%
  \BibitemOpen
  \bibfield  {author} {\bibinfo {author} {\bibfnamefont {G.}~\bibnamefont
  {{Benevento}}}, \bibinfo {author} {\bibfnamefont {W.}~\bibnamefont {{Hu}}},\
  and\ \bibinfo {author} {\bibfnamefont {M.}~\bibnamefont {{Raveri}}},\
  }\bibfield  {title} {\bibinfo {title} {{Can late dark energy transitions
  raise the Hubble constant?}},\ }\href
  {https://doi.org/10.1103/PhysRevD.101.103517} {\bibfield  {journal} {\bibinfo
   {journal} {\prd}\ }\textbf {\bibinfo {volume} {101}},\ \bibinfo {eid}
  {103517} (\bibinfo {year} {2020})}\BibitemShut {NoStop}%
\bibitem [{\citenamefont {{Hern{\'a}ndez-Almada}}\ \emph
  {et~al.}(2020)\citenamefont {{Hern{\'a}ndez-Almada}}, \citenamefont {{Leon}},
  \citenamefont {{Maga{\~n}a}}, \citenamefont {{Garc{\'\i}a-Aspeitia}},\ and\
  \citenamefont {{Motta}}}]{Hernandez-Almada:2020uyr}%
  \BibitemOpen
  \bibfield  {author} {\bibinfo {author} {\bibfnamefont {A.}~\bibnamefont
  {{Hern{\'a}ndez-Almada}}}, \bibinfo {author} {\bibfnamefont {G.}~\bibnamefont
  {{Leon}}}, \bibinfo {author} {\bibfnamefont {J.}~\bibnamefont
  {{Maga{\~n}a}}}, \bibinfo {author} {\bibfnamefont {M.~A.}\ \bibnamefont
  {{Garc{\'\i}a-Aspeitia}}},\ and\ \bibinfo {author} {\bibfnamefont
  {V.}~\bibnamefont {{Motta}}},\ }\bibfield  {title} {\bibinfo {title}
  {{Generalized emergent dark energy: observational Hubble data constraints and
  stability analysis}},\ }\href {https://doi.org/10.1093/mnras/staa2052}
  {\bibfield  {journal} {\bibinfo  {journal} {\mnras}\ }\textbf {\bibinfo
  {volume} {497}},\ \bibinfo {pages} {1590} (\bibinfo {year}
  {2020})}\BibitemShut {NoStop}%
\bibitem [{\citenamefont {Barker}\ \emph {et~al.}(2020)\citenamefont {Barker},
  \citenamefont {Lasenby}, \citenamefont {Hobson},\ and\ \citenamefont
  {Handley}}]{Barker:2020gcp}%
  \BibitemOpen
  \bibfield  {author} {\bibinfo {author} {\bibfnamefont {W.~E.~V.}\
  \bibnamefont {Barker}}, \bibinfo {author} {\bibfnamefont {A.~N.}\
  \bibnamefont {Lasenby}}, \bibinfo {author} {\bibfnamefont {M.~P.}\
  \bibnamefont {Hobson}},\ and\ \bibinfo {author} {\bibfnamefont {W.~J.}\
  \bibnamefont {Handley}},\ }\bibfield  {title} {\bibinfo {title} {{Systematic
  study of background cosmology in unitary Poincaré gauge theories with
  application to emergent dark radiation and $H_0$ tension}},\ }\href
  {https://doi.org/10.1103/PhysRevD.102.024048} {\bibfield  {journal} {\bibinfo
   {journal} {prd}\ }\textbf {\bibinfo {volume} {102}},\ \bibinfo {pages}
  {024048} (\bibinfo {year} {2020})}\BibitemShut {NoStop}%
\bibitem [{\citenamefont {{Jedamzik}}\ and\ \citenamefont
  {{Pogosian}}(2020)}]{2020PhRvL.125r1302J}%
  \BibitemOpen
  \bibfield  {author} {\bibinfo {author} {\bibfnamefont {K.}~\bibnamefont
  {{Jedamzik}}}\ and\ \bibinfo {author} {\bibfnamefont {L.}~\bibnamefont
  {{Pogosian}}},\ }\bibfield  {title} {\bibinfo {title} {{Relieving the Hubble
  Tension with Primordial Magnetic Fields}},\ }\href
  {https://doi.org/10.1103/PhysRevLett.125.181302} {\bibfield  {journal}
  {\bibinfo  {journal} {\prl}\ }\textbf {\bibinfo {volume} {125}},\ \bibinfo
  {eid} {181302} (\bibinfo {year} {2020})}\BibitemShut {NoStop}%
\bibitem [{\citenamefont {{Ballardini}}\ \emph {et~al.}(2020)\citenamefont
  {{Ballardini}}, \citenamefont {{Braglia}}, \citenamefont {{Finelli}},
  \citenamefont {{Paoletti}}, \citenamefont {{Starobinsky}},\ and\
  \citenamefont {{Umilt{\`a}}}}]{2020JCAP...10..044B}%
  \BibitemOpen
  \bibfield  {author} {\bibinfo {author} {\bibfnamefont {M.}~\bibnamefont
  {{Ballardini}}}, \bibinfo {author} {\bibfnamefont {M.}~\bibnamefont
  {{Braglia}}}, \bibinfo {author} {\bibfnamefont {F.}~\bibnamefont
  {{Finelli}}}, \bibinfo {author} {\bibfnamefont {D.}~\bibnamefont
  {{Paoletti}}}, \bibinfo {author} {\bibfnamefont {A.~A.}\ \bibnamefont
  {{Starobinsky}}},\ and\ \bibinfo {author} {\bibfnamefont {C.}~\bibnamefont
  {{Umilt{\`a}}}},\ }\bibfield  {title} {\bibinfo {title} {{Scalar-tensor
  theories of gravity, neutrino physics, and the H$_{0}$ tension}},\ }\href
  {https://doi.org/10.1088/1475-7516/2020/10/044} {\bibfield  {journal}
  {\bibinfo  {journal} {\jcap}\ }\textbf {\bibinfo {volume} {2020}},\ \bibinfo
  {eid} {044} (\bibinfo {year} {2020})}\BibitemShut {NoStop}%
\bibitem [{\citenamefont {{Banerjee}}\ \emph {et~al.}()\citenamefont
  {{Banerjee}}, \citenamefont {{Cai}}, \citenamefont {{Heisenberg}},
  \citenamefont {{Colg{\'a}in}}, \citenamefont {{Sheikh-Jabbari}},\ and\
  \citenamefont {{Yang}}}]{Banerjee:2020xcn}%
  \BibitemOpen
  \bibfield  {author} {\bibinfo {author} {\bibfnamefont {A.}~\bibnamefont
  {{Banerjee}}}, \bibinfo {author} {\bibfnamefont {H.}~\bibnamefont {{Cai}}},
  \bibinfo {author} {\bibfnamefont {L.}~\bibnamefont {{Heisenberg}}}, \bibinfo
  {author} {\bibfnamefont {E.~{\'O}.}\ \bibnamefont {{Colg{\'a}in}}}, \bibinfo
  {author} {\bibfnamefont {M.~M.}\ \bibnamefont {{Sheikh-Jabbari}}},\ and\
  \bibinfo {author} {\bibfnamefont {T.}~\bibnamefont {{Yang}}},\ }\href@noop {}
  {\bibinfo {title} {{Hubble Sinks In The Low-Redshift Swampland}}},\ \Eprint
  {https://arxiv.org/abs/2006.00244} {arXiv:2006.00244} \BibitemShut {NoStop}%
\bibitem [{\citenamefont {{Sol{\`a} Peracaula}}\ \emph
  {et~al.}(2020)\citenamefont {{Sol{\`a} Peracaula}}, \citenamefont
  {{G{\'o}mez-Valent}}, \citenamefont {{de Cruz P{\'e}rez}},\ and\
  \citenamefont {{Moreno-Pulido}}}]{Sola:2020lba}%
  \BibitemOpen
  \bibfield  {author} {\bibinfo {author} {\bibfnamefont {J.}~\bibnamefont
  {{Sol{\`a} Peracaula}}}, \bibinfo {author} {\bibfnamefont {A.}~\bibnamefont
  {{G{\'o}mez-Valent}}}, \bibinfo {author} {\bibfnamefont {J.}~\bibnamefont
  {{de Cruz P{\'e}rez}}},\ and\ \bibinfo {author} {\bibfnamefont
  {C.}~\bibnamefont {{Moreno-Pulido}}},\ }\bibfield  {title} {\bibinfo {title}
  {{Brans-Dicke cosmology with a {\ensuremath{\Lambda}}-term: a possible
  solution to {\ensuremath{\Lambda}}CDM tensions}},\ }\href
  {https://doi.org/10.1088/1361-6382/abbc43} {\bibfield  {journal} {\bibinfo
  {journal} {Classical and Quantum Gravity}\ }\textbf {\bibinfo {volume}
  {37}},\ \bibinfo {eid} {245003} (\bibinfo {year} {2020})}\BibitemShut
  {NoStop}%
\bibitem [{\citenamefont {{Sekiguchi}}\ and\ \citenamefont
  {{Takahashi}}(2020)}]{2020arXiv200703381S}%
  \BibitemOpen
  \bibfield  {author} {\bibinfo {author} {\bibfnamefont {T.}~\bibnamefont
  {{Sekiguchi}}}\ and\ \bibinfo {author} {\bibfnamefont {T.}~\bibnamefont
  {{Takahashi}}},\ }\href@noop {} {\bibinfo {title} {{Early recombination as a
  solution to the $H_0$ tension}}} (\bibinfo {year} {2020}),\ \Eprint
  {https://arxiv.org/abs/2007.03381} {arXiv:2007.03381} \BibitemShut {NoStop}%
\bibitem [{\citenamefont {Benaoum}\ \emph {et~al.}()\citenamefont {Benaoum},
  \citenamefont {Yang}, \citenamefont {Pan},\ and\ \citenamefont
  {Di~Valentino}}]{Benaoum:2020qsi}%
  \BibitemOpen
  \bibfield  {author} {\bibinfo {author} {\bibfnamefont {H.}~\bibnamefont
  {Benaoum}}, \bibinfo {author} {\bibfnamefont {W.}~\bibnamefont {Yang}},
  \bibinfo {author} {\bibfnamefont {S.}~\bibnamefont {Pan}},\ and\ \bibinfo
  {author} {\bibfnamefont {E.}~\bibnamefont {Di~Valentino}},\ }\href@noop {}
  {\bibinfo {title} {{Modified Emergent Dark Energy and its Astronomical
  Constraints}}},\ \Eprint {https://arxiv.org/abs/2008.09098}
  {arXiv:2008.09098} \BibitemShut {NoStop}%
\bibitem [{\citenamefont {{Aghanim}}\ \emph {et~al.}(2020)\citenamefont
  {{Aghanim}}, \citenamefont {{Akrami}}, \citenamefont {{Ashdown}},
  \citenamefont {{Aumont}}, \citenamefont {{Baccigalupi}}, \citenamefont
  {{Ballardini}}, \citenamefont {{Banday}}, \citenamefont {{Barreiro}},
  \citenamefont {{Bartolo}} \emph {et~al.}}]{2020A&A...641A...6P}%
  \BibitemOpen
  \bibfield  {author} {\bibinfo {author} {\bibfnamefont {N.}~\bibnamefont
  {{Aghanim}}}, \bibinfo {author} {\bibfnamefont {Y.}~\bibnamefont {{Akrami}}},
  \bibinfo {author} {\bibfnamefont {M.}~\bibnamefont {{Ashdown}}}, \bibinfo
  {author} {\bibfnamefont {J.}~\bibnamefont {{Aumont}}}, \bibinfo {author}
  {\bibfnamefont {C.}~\bibnamefont {{Baccigalupi}}}, \bibinfo {author}
  {\bibfnamefont {M.}~\bibnamefont {{Ballardini}}}, \bibinfo {author}
  {\bibfnamefont {A.~J.}\ \bibnamefont {{Banday}}}, \bibinfo {author}
  {\bibfnamefont {R.~B.}\ \bibnamefont {{Barreiro}}}, \bibinfo {author}
  {\bibfnamefont {N.}~\bibnamefont {{Bartolo}}}, \emph {et~al.} (\bibinfo
  {collaboration} {Planck Collaboration}),\ }\bibfield  {title} {\bibinfo
  {title} {{Planck 2018 results. VI. Cosmological parameters}},\ }\href
  {https://doi.org/10.1051/0004-6361/201833910} {\bibfield  {journal} {\bibinfo
   {journal} {\aap}\ }\textbf {\bibinfo {volume} {641}},\ \bibinfo {eid} {A6}
  (\bibinfo {year} {2020})}\BibitemShut {NoStop}%
\bibitem [{\citenamefont {Knox}\ and\ \citenamefont
  {Millea}(2020)}]{Knox:2019rjx}%
  \BibitemOpen
  \bibfield  {author} {\bibinfo {author} {\bibfnamefont {L.}~\bibnamefont
  {Knox}}\ and\ \bibinfo {author} {\bibfnamefont {M.}~\bibnamefont {Millea}},\
  }\bibfield  {title} {\bibinfo {title} {{Hubble constant hunter's guide}},\
  }\href {https://doi.org/10.1103/PhysRevD.101.043533} {\bibfield  {journal}
  {\bibinfo  {journal} {\prd}\ }\textbf {\bibinfo {volume} {101}},\ \bibinfo
  {pages} {043533} (\bibinfo {year} {2020})}\BibitemShut {NoStop}%
\bibitem [{\citenamefont {{Komatsu}}\ \emph {et~al.}(2011)\citenamefont
  {{Komatsu}}, \citenamefont {{Smith}}, \citenamefont {{Dunkley}},
  \citenamefont {{Bennett}}, \citenamefont {{Gold}}, \citenamefont {{Hinshaw}},
  \citenamefont {{Jarosik}}, \citenamefont {{Larson}}, \citenamefont {{Nolta}},
  \citenamefont {{Page}} \emph {et~al.}}]{2011ApJS..192...18K}%
  \BibitemOpen
  \bibfield  {author} {\bibinfo {author} {\bibfnamefont {E.}~\bibnamefont
  {{Komatsu}}}, \bibinfo {author} {\bibfnamefont {K.~M.}\ \bibnamefont
  {{Smith}}}, \bibinfo {author} {\bibfnamefont {J.}~\bibnamefont {{Dunkley}}},
  \bibinfo {author} {\bibfnamefont {C.~L.}\ \bibnamefont {{Bennett}}}, \bibinfo
  {author} {\bibfnamefont {B.}~\bibnamefont {{Gold}}}, \bibinfo {author}
  {\bibfnamefont {G.}~\bibnamefont {{Hinshaw}}}, \bibinfo {author}
  {\bibfnamefont {N.}~\bibnamefont {{Jarosik}}}, \bibinfo {author}
  {\bibfnamefont {D.}~\bibnamefont {{Larson}}}, \bibinfo {author}
  {\bibfnamefont {M.~R.}\ \bibnamefont {{Nolta}}}, \bibinfo {author}
  {\bibfnamefont {L.}~\bibnamefont {{Page}}}, \emph {et~al.},\ }\bibfield
  {title} {\bibinfo {title} {{Seven-year Wilkinson Microwave Anisotropy Probe
  (WMAP) Observations: Cosmological Interpretation}},\ }\href
  {https://doi.org/10.1088/0067-0049/192/2/18} {\bibfield  {journal} {\bibinfo
  {journal} {\apjs}\ }\textbf {\bibinfo {volume} {192}},\ \bibinfo {eid} {18}
  (\bibinfo {year} {2011})}\BibitemShut {NoStop}%
\bibitem [{\citenamefont {Chevallier}\ and\ \citenamefont
  {Polarski}(2001)}]{Chevallier:2000qy}%
  \BibitemOpen
  \bibfield  {author} {\bibinfo {author} {\bibfnamefont {M.}~\bibnamefont
  {Chevallier}}\ and\ \bibinfo {author} {\bibfnamefont {D.}~\bibnamefont
  {Polarski}},\ }\bibfield  {title} {\bibinfo {title} {{Accelerating universes
  with scaling dark matter}},\ }\href
  {https://doi.org/10.1142/S0218271801000822} {\bibfield  {journal} {\bibinfo
  {journal} {Int. J. Mod. Phys. D}\ }\textbf {\bibinfo {volume} {10}},\
  \bibinfo {pages} {213} (\bibinfo {year} {2001})}\BibitemShut {NoStop}%
\bibitem [{\citenamefont {Linder}(2003)}]{Linder:2002et}%
  \BibitemOpen
  \bibfield  {author} {\bibinfo {author} {\bibfnamefont {E.~V.}\ \bibnamefont
  {Linder}},\ }\bibfield  {title} {\bibinfo {title} {{Exploring the expansion
  history of the universe}},\ }\href
  {https://doi.org/10.1103/PhysRevLett.90.091301} {\bibfield  {journal}
  {\bibinfo  {journal} {Phys. Rev. Lett.}\ }\textbf {\bibinfo {volume} {90}},\
  \bibinfo {pages} {091301} (\bibinfo {year} {2003})}\BibitemShut {NoStop}%
\bibitem [{\citenamefont {{Huang}}\ \emph {et~al.}(2020)\citenamefont
  {{Huang}}, \citenamefont {{Riess}}, \citenamefont {{Yuan}}, \citenamefont
  {{Macri}}, \citenamefont {{Zakamska}}, \citenamefont {{Casertano}},
  \citenamefont {{Whitelock}}, \citenamefont {{Hoffmann}}, \citenamefont
  {{Filippenko}},\ and\ \citenamefont {{Scolnic}}}]{2020ApJ...889....5H}%
  \BibitemOpen
  \bibfield  {author} {\bibinfo {author} {\bibfnamefont {C.~D.}\ \bibnamefont
  {{Huang}}}, \bibinfo {author} {\bibfnamefont {A.~G.}\ \bibnamefont
  {{Riess}}}, \bibinfo {author} {\bibfnamefont {W.}~\bibnamefont {{Yuan}}},
  \bibinfo {author} {\bibfnamefont {L.~M.}\ \bibnamefont {{Macri}}}, \bibinfo
  {author} {\bibfnamefont {N.~L.}\ \bibnamefont {{Zakamska}}}, \bibinfo
  {author} {\bibfnamefont {S.}~\bibnamefont {{Casertano}}}, \bibinfo {author}
  {\bibfnamefont {P.~A.}\ \bibnamefont {{Whitelock}}}, \bibinfo {author}
  {\bibfnamefont {S.~L.}\ \bibnamefont {{Hoffmann}}}, \bibinfo {author}
  {\bibfnamefont {A.~V.}\ \bibnamefont {{Filippenko}}},\ and\ \bibinfo {author}
  {\bibfnamefont {D.}~\bibnamefont {{Scolnic}}},\ }\bibfield  {title} {\bibinfo
  {title} {{Hubble Space Telescope Observations of Mira Variables in the SN Ia
  Host NGC 1559: An Alternative Candle to Measure the Hubble Constant}},\
  }\href {https://doi.org/10.3847/1538-4357/ab5dbd} {\bibfield  {journal}
  {\bibinfo  {journal} {\apj}\ }\textbf {\bibinfo {volume} {889}},\ \bibinfo
  {eid} {5} (\bibinfo {year} {2020})}\BibitemShut {NoStop}%
\bibitem [{\citenamefont {Riess}\ \emph {et~al.}(2019)\citenamefont {Riess},
  \citenamefont {Casertano}, \citenamefont {Yuan}, \citenamefont {Macri},\ and\
  \citenamefont {Scolnic}}]{Riess:2019cxk}%
  \BibitemOpen
  \bibfield  {author} {\bibinfo {author} {\bibfnamefont {A.~G.}\ \bibnamefont
  {Riess}}, \bibinfo {author} {\bibfnamefont {S.}~\bibnamefont {Casertano}},
  \bibinfo {author} {\bibfnamefont {W.}~\bibnamefont {Yuan}}, \bibinfo {author}
  {\bibfnamefont {L.~M.}\ \bibnamefont {Macri}},\ and\ \bibinfo {author}
  {\bibfnamefont {D.}~\bibnamefont {Scolnic}},\ }\bibfield  {title} {\bibinfo
  {title} {{Large Magellanic Cloud Cepheid Standards Provide a 1\% Foundation
  for the Determination of the Hubble Constant and Stronger Evidence for
  Physics beyond $\Lambda$CDM}},\ }\href
  {https://doi.org/10.3847/1538-4357/ab1422} {\bibfield  {journal} {\bibinfo
  {journal} {Astrophys. J.}\ }\textbf {\bibinfo {volume} {876}},\ \bibinfo
  {pages} {85} (\bibinfo {year} {2019})}\BibitemShut {NoStop}%
\bibitem [{\citenamefont {{Freedman}}\ \emph {et~al.}(2019)\citenamefont
  {{Freedman}}, \citenamefont {{Madore}}, \citenamefont {{Hatt}}, \citenamefont
  {{Hoyt}}, \citenamefont {{Jang}}, \citenamefont {{Beaton}}, \citenamefont
  {{Burns}}, \citenamefont {{Lee}}, \citenamefont {{Monson}}, \citenamefont
  {{Neeley}}, \citenamefont {{Phillips}}, \citenamefont {{Rich}},\ and\
  \citenamefont {{Seibert}}}]{2019ApJ...882...34F}%
  \BibitemOpen
  \bibfield  {author} {\bibinfo {author} {\bibfnamefont {W.~L.}\ \bibnamefont
  {{Freedman}}}, \bibinfo {author} {\bibfnamefont {B.~F.}\ \bibnamefont
  {{Madore}}}, \bibinfo {author} {\bibfnamefont {D.}~\bibnamefont {{Hatt}}},
  \bibinfo {author} {\bibfnamefont {T.~J.}\ \bibnamefont {{Hoyt}}}, \bibinfo
  {author} {\bibfnamefont {I.~S.}\ \bibnamefont {{Jang}}}, \bibinfo {author}
  {\bibfnamefont {R.~L.}\ \bibnamefont {{Beaton}}}, \bibinfo {author}
  {\bibfnamefont {C.~R.}\ \bibnamefont {{Burns}}}, \bibinfo {author}
  {\bibfnamefont {M.~G.}\ \bibnamefont {{Lee}}}, \bibinfo {author}
  {\bibfnamefont {A.~J.}\ \bibnamefont {{Monson}}}, \bibinfo {author}
  {\bibfnamefont {J.~R.}\ \bibnamefont {{Neeley}}}, \bibinfo {author}
  {\bibfnamefont {M.~M.}\ \bibnamefont {{Phillips}}}, \bibinfo {author}
  {\bibfnamefont {J.~A.}\ \bibnamefont {{Rich}}},\ and\ \bibinfo {author}
  {\bibfnamefont {M.}~\bibnamefont {{Seibert}}},\ }\bibfield  {title} {\bibinfo
  {title} {{The Carnegie-Chicago Hubble Program. VIII. An Independent
  Determination of the Hubble Constant Based on the Tip of the Red Giant
  Branch}},\ }\href {https://doi.org/10.3847/1538-4357/ab2f73} {\bibfield
  {journal} {\bibinfo  {journal} {\apj}\ }\textbf {\bibinfo {volume} {882}},\
  \bibinfo {eid} {34} (\bibinfo {year} {2019})}\BibitemShut {NoStop}%
\bibitem [{\citenamefont {{Skilling}}(2004)}]{2004AIPC..735..395S}%
  \BibitemOpen
  \bibfield  {author} {\bibinfo {author} {\bibfnamefont {J.}~\bibnamefont
  {{Skilling}}},\ }\bibfield  {title} {\bibinfo {title} {{Nested Sampling}},\
  }in\ \href {https://doi.org/10.1063/1.1835238} {\emph {\bibinfo {booktitle}
  {American Institute of Physics Conference Series}}},\ \bibinfo {series}
  {American Institute of Physics Conference Series}, Vol.\ \bibinfo {volume}
  {735},\ \bibinfo {editor} {edited by\ \bibinfo {editor} {\bibfnamefont
  {R.}~\bibnamefont {{Fischer}}}, \bibinfo {editor} {\bibfnamefont
  {R.}~\bibnamefont {{Preuss}}},\ and\ \bibinfo {editor} {\bibfnamefont
  {U.~V.}\ \bibnamefont {{Toussaint}}}}\ (\bibinfo  {publisher} {AIP Publishing
  LLC},\ \bibinfo {year} {2004})\ pp.\ \bibinfo {pages} {395--405}\BibitemShut
  {NoStop}%
\bibitem [{\citenamefont {{Feroz}}\ \emph {et~al.}(2009)\citenamefont
  {{Feroz}}, \citenamefont {{Hobson}},\ and\ \citenamefont
  {{Bridges}}}]{2009MNRAS.398.1601F}%
  \BibitemOpen
  \bibfield  {author} {\bibinfo {author} {\bibfnamefont {F.}~\bibnamefont
  {{Feroz}}}, \bibinfo {author} {\bibfnamefont {M.~P.}\ \bibnamefont
  {{Hobson}}},\ and\ \bibinfo {author} {\bibfnamefont {M.}~\bibnamefont
  {{Bridges}}},\ }\bibfield  {title} {\bibinfo {title} {{MULTINEST: an
  efficient and robust Bayesian inference tool for cosmology and particle
  physics}},\ }\href {https://doi.org/10.1111/j.1365-2966.2009.14548.x}
  {\bibfield  {journal} {\bibinfo  {journal} {\mnras}\ }\textbf {\bibinfo
  {volume} {398}},\ \bibinfo {pages} {1601} (\bibinfo {year}
  {2009})}\BibitemShut {NoStop}%
\bibitem [{\citenamefont {{Buchner}}\ \emph {et~al.}(2014)\citenamefont
  {{Buchner}}, \citenamefont {{Georgakakis}}, \citenamefont {{Nandra}},
  \citenamefont {{Hsu}}, \citenamefont {{Rangel}}, \citenamefont {{Brightman}},
  \citenamefont {{Merloni}}, \citenamefont {{Salvato}}, \citenamefont
  {{Donley}},\ and\ \citenamefont {{Kocevski}}}]{2014A&A...564A.125B}%
  \BibitemOpen
  \bibfield  {author} {\bibinfo {author} {\bibfnamefont {J.}~\bibnamefont
  {{Buchner}}}, \bibinfo {author} {\bibfnamefont {A.}~\bibnamefont
  {{Georgakakis}}}, \bibinfo {author} {\bibfnamefont {K.}~\bibnamefont
  {{Nandra}}}, \bibinfo {author} {\bibfnamefont {L.}~\bibnamefont {{Hsu}}},
  \bibinfo {author} {\bibfnamefont {C.}~\bibnamefont {{Rangel}}}, \bibinfo
  {author} {\bibfnamefont {M.}~\bibnamefont {{Brightman}}}, \bibinfo {author}
  {\bibfnamefont {A.}~\bibnamefont {{Merloni}}}, \bibinfo {author}
  {\bibfnamefont {M.}~\bibnamefont {{Salvato}}}, \bibinfo {author}
  {\bibfnamefont {J.}~\bibnamefont {{Donley}}},\ and\ \bibinfo {author}
  {\bibfnamefont {D.}~\bibnamefont {{Kocevski}}},\ }\bibfield  {title}
  {\bibinfo {title} {{X-ray spectral modelling of the AGN obscuring region in
  the CDFS: Bayesian model selection and catalogue}},\ }\href
  {https://doi.org/10.1051/0004-6361/201322971} {\bibfield  {journal} {\bibinfo
   {journal} {\aap}\ }\textbf {\bibinfo {volume} {564}},\ \bibinfo {eid} {A125}
  (\bibinfo {year} {2014})}\BibitemShut {NoStop}%
\bibitem [{\citenamefont {{Scolnic}}\ \emph {et~al.}(2018)\citenamefont
  {{Scolnic}}, \citenamefont {{Jones}}, \citenamefont {{Rest}}, \citenamefont
  {{Pan}}, \citenamefont {{Chornock}}, \citenamefont {{Foley}}, \citenamefont
  {{Huber}}, \citenamefont {{Kessler}}, \citenamefont {{Narayan}},
  \citenamefont {{Riess}} \emph {et~al.}}]{2018ApJ...859..101S}%
  \BibitemOpen
  \bibfield  {author} {\bibinfo {author} {\bibfnamefont {D.~M.}\ \bibnamefont
  {{Scolnic}}}, \bibinfo {author} {\bibfnamefont {D.~O.}\ \bibnamefont
  {{Jones}}}, \bibinfo {author} {\bibfnamefont {A.}~\bibnamefont {{Rest}}},
  \bibinfo {author} {\bibfnamefont {Y.~C.}\ \bibnamefont {{Pan}}}, \bibinfo
  {author} {\bibfnamefont {R.}~\bibnamefont {{Chornock}}}, \bibinfo {author}
  {\bibfnamefont {R.~J.}\ \bibnamefont {{Foley}}}, \bibinfo {author}
  {\bibfnamefont {M.~E.}\ \bibnamefont {{Huber}}}, \bibinfo {author}
  {\bibfnamefont {R.}~\bibnamefont {{Kessler}}}, \bibinfo {author}
  {\bibfnamefont {G.}~\bibnamefont {{Narayan}}}, \bibinfo {author}
  {\bibfnamefont {A.~G.}\ \bibnamefont {{Riess}}}, \emph {et~al.},\ }\bibfield
  {title} {\bibinfo {title} {{The Complete Light-curve Sample of
  Spectroscopically Confirmed SNe Ia from Pan-STARRS1 and Cosmological
  Constraints from the Combined Pantheon Sample}},\ }\href
  {https://doi.org/10.3847/1538-4357/aab9bb} {\bibfield  {journal} {\bibinfo
  {journal} {\apj}\ }\textbf {\bibinfo {volume} {859}},\ \bibinfo {eid} {101}
  (\bibinfo {year} {2018})}\BibitemShut {NoStop}%
\bibitem [{\citenamefont {{Alam}}\ \emph {et~al.}(2017)\citenamefont {{Alam}},
  \citenamefont {{Ata}}, \citenamefont {{Bailey}}, \citenamefont {{Beutler}},
  \citenamefont {{Bizyaev}}, \citenamefont {{Blazek}}, \citenamefont
  {{Bolton}}, \citenamefont {{Brownstein}}, \citenamefont {{Burden}},
  \citenamefont {{Chuang}} \emph {et~al.}}]{2017MNRAS.470.2617A}%
  \BibitemOpen
  \bibfield  {author} {\bibinfo {author} {\bibfnamefont {S.}~\bibnamefont
  {{Alam}}}, \bibinfo {author} {\bibfnamefont {M.}~\bibnamefont {{Ata}}},
  \bibinfo {author} {\bibfnamefont {S.}~\bibnamefont {{Bailey}}}, \bibinfo
  {author} {\bibfnamefont {F.}~\bibnamefont {{Beutler}}}, \bibinfo {author}
  {\bibfnamefont {D.}~\bibnamefont {{Bizyaev}}}, \bibinfo {author}
  {\bibfnamefont {J.~A.}\ \bibnamefont {{Blazek}}}, \bibinfo {author}
  {\bibfnamefont {A.~S.}\ \bibnamefont {{Bolton}}}, \bibinfo {author}
  {\bibfnamefont {J.~R.}\ \bibnamefont {{Brownstein}}}, \bibinfo {author}
  {\bibfnamefont {A.}~\bibnamefont {{Burden}}}, \bibinfo {author}
  {\bibfnamefont {C.-H.}\ \bibnamefont {{Chuang}}}, \emph {et~al.},\ }\bibfield
   {title} {\bibinfo {title} {{The clustering of galaxies in the completed
  SDSS-III Baryon Oscillation Spectroscopic Survey: cosmological analysis of
  the DR12 galaxy sample}},\ }\href {https://doi.org/10.1093/mnras/stx721}
  {\bibfield  {journal} {\bibinfo  {journal} {\mnras}\ }\textbf {\bibinfo
  {volume} {470}},\ \bibinfo {pages} {2617} (\bibinfo {year}
  {2017})}\BibitemShut {NoStop}%
\bibitem [{\citenamefont {{Zhao}}\ \emph {et~al.}(2017)\citenamefont {{Zhao}},
  \citenamefont {{Wang}}, \citenamefont {{Saito}}, \citenamefont {{Wang}},
  \citenamefont {{Ross}}, \citenamefont {{Beutler}}, \citenamefont {{Grieb}},
  \citenamefont {{Chuang}}, \citenamefont {{Kitaura}}, \citenamefont
  {{Rodriguez-Torres}} \emph {et~al.}}]{2017MNRAS.466..762Z}%
  \BibitemOpen
  \bibfield  {author} {\bibinfo {author} {\bibfnamefont {G.-B.}\ \bibnamefont
  {{Zhao}}}, \bibinfo {author} {\bibfnamefont {Y.}~\bibnamefont {{Wang}}},
  \bibinfo {author} {\bibfnamefont {S.}~\bibnamefont {{Saito}}}, \bibinfo
  {author} {\bibfnamefont {D.}~\bibnamefont {{Wang}}}, \bibinfo {author}
  {\bibfnamefont {A.~J.}\ \bibnamefont {{Ross}}}, \bibinfo {author}
  {\bibfnamefont {F.}~\bibnamefont {{Beutler}}}, \bibinfo {author}
  {\bibfnamefont {J.~N.}\ \bibnamefont {{Grieb}}}, \bibinfo {author}
  {\bibfnamefont {C.-H.}\ \bibnamefont {{Chuang}}}, \bibinfo {author}
  {\bibfnamefont {F.-S.}\ \bibnamefont {{Kitaura}}}, \bibinfo {author}
  {\bibfnamefont {S.}~\bibnamefont {{Rodriguez-Torres}}}, \emph {et~al.},\
  }\bibfield  {title} {\bibinfo {title} {{The clustering of galaxies in the
  completed SDSS-III Baryon Oscillation Spectroscopic Survey: tomographic BAO
  analysis of DR12 combined sample in Fourier space}},\ }\href
  {https://doi.org/10.1093/mnras/stw3199} {\bibfield  {journal} {\bibinfo
  {journal} {\mnras}\ }\textbf {\bibinfo {volume} {466}},\ \bibinfo {pages}
  {762} (\bibinfo {year} {2017})}\BibitemShut {NoStop}%
\bibitem [{\citenamefont {{Astropy Collaboration}}\ \emph
  {et~al.}(2018)\citenamefont {{Astropy Collaboration}}, \citenamefont
  {{Price-Whelan}}, \citenamefont {{Sip{\H o}cz}}, \citenamefont
  {{G{\"u}nther}}, \citenamefont {{Lim}}, \citenamefont {{Crawford}},
  \citenamefont {{Conseil}}, \citenamefont {{Shupe}}, \citenamefont {{Craig}},
  \citenamefont {{Dencheva}} \emph {et~al.}}]{2018AJ....156..123A}%
  \BibitemOpen
  \bibfield  {author} {\bibinfo {author} {\bibnamefont {{Astropy
  Collaboration}}}, \bibinfo {author} {\bibfnamefont {A.~M.}\ \bibnamefont
  {{Price-Whelan}}}, \bibinfo {author} {\bibfnamefont {B.~M.}\ \bibnamefont
  {{Sip{\H o}cz}}}, \bibinfo {author} {\bibfnamefont {H.~M.}\ \bibnamefont
  {{G{\"u}nther}}}, \bibinfo {author} {\bibfnamefont {P.~L.}\ \bibnamefont
  {{Lim}}}, \bibinfo {author} {\bibfnamefont {S.~M.}\ \bibnamefont
  {{Crawford}}}, \bibinfo {author} {\bibfnamefont {S.}~\bibnamefont
  {{Conseil}}}, \bibinfo {author} {\bibfnamefont {D.~L.}\ \bibnamefont
  {{Shupe}}}, \bibinfo {author} {\bibfnamefont {M.~W.}\ \bibnamefont
  {{Craig}}}, \bibinfo {author} {\bibfnamefont {N.}~\bibnamefont {{Dencheva}}},
  \emph {et~al.},\ }\bibfield  {title} {\bibinfo {title} {{The Astropy Project:
  Building an Open-science Project and Status of the v2.0 Core Package}},\
  }\href {https://doi.org/10.3847/1538-3881/aabc4f} {\bibfield  {journal}
  {\bibinfo  {journal} {\aj}\ }\textbf {\bibinfo {volume} {156}},\ \bibinfo
  {eid} {123} (\bibinfo {year} {2018})}\BibitemShut {NoStop}%
\bibitem [{\citenamefont {{Astropy Collaboration}}\ \emph
  {et~al.}(2013)\citenamefont {{Astropy Collaboration}}, \citenamefont
  {{Robitaille}}, \citenamefont {{Tollerud}}, \citenamefont {{Greenfield}},
  \citenamefont {{Droettboom}}, \citenamefont {{Bray}}, \citenamefont
  {{Aldcroft}}, \citenamefont {{Davis}}, \citenamefont {{Ginsburg}},
  \citenamefont {{Price-Whelan}} \emph {et~al.}}]{2013A&A...558A..33A}%
  \BibitemOpen
  \bibfield  {author} {\bibinfo {author} {\bibnamefont {{Astropy
  Collaboration}}}, \bibinfo {author} {\bibfnamefont {T.~P.}\ \bibnamefont
  {{Robitaille}}}, \bibinfo {author} {\bibfnamefont {E.~J.}\ \bibnamefont
  {{Tollerud}}}, \bibinfo {author} {\bibfnamefont {P.}~\bibnamefont
  {{Greenfield}}}, \bibinfo {author} {\bibfnamefont {M.}~\bibnamefont
  {{Droettboom}}}, \bibinfo {author} {\bibfnamefont {E.}~\bibnamefont
  {{Bray}}}, \bibinfo {author} {\bibfnamefont {T.}~\bibnamefont {{Aldcroft}}},
  \bibinfo {author} {\bibfnamefont {M.}~\bibnamefont {{Davis}}}, \bibinfo
  {author} {\bibfnamefont {A.}~\bibnamefont {{Ginsburg}}}, \bibinfo {author}
  {\bibfnamefont {A.~M.}\ \bibnamefont {{Price-Whelan}}}, \emph {et~al.},\
  }\bibfield  {title} {\bibinfo {title} {{Astropy: A community Python package
  for astronomy}},\ }\href {https://doi.org/10.1051/0004-6361/201322068}
  {\bibfield  {journal} {\bibinfo  {journal} {\aap}\ }\textbf {\bibinfo
  {volume} {558}},\ \bibinfo {eid} {A33} (\bibinfo {year} {2013})}\BibitemShut
  {NoStop}%
\bibitem [{\citenamefont {{Hinton}}(2016)}]{Hinton2016}%
  \BibitemOpen
  \bibfield  {author} {\bibinfo {author} {\bibfnamefont {S.~R.}\ \bibnamefont
  {{Hinton}}},\ }\bibfield  {title} {\bibinfo {title} {{ChainConsumer}},\
  }\href {https://doi.org/10.21105/joss.00045} {\bibfield  {journal} {\bibinfo
  {journal} {J. Open Source Softw.}\ }\textbf {\bibinfo {volume} {1}},\
  \bibinfo {eid} {00045} (\bibinfo {year} {2016})}\BibitemShut {NoStop}%
\bibitem [{\citenamefont {Hunter}(2007)}]{Hunter:2007}%
  \BibitemOpen
  \bibfield  {author} {\bibinfo {author} {\bibfnamefont {J.~D.}\ \bibnamefont
  {Hunter}},\ }\bibfield  {title} {\bibinfo {title} {Matplotlib: A 2d graphics
  environment},\ }\href {https://doi.org/10.1109/MCSE.2007.55} {\bibfield
  {journal} {\bibinfo  {journal} {Comp. Sci. Eng.}\ }\textbf {\bibinfo {volume}
  {9}},\ \bibinfo {pages} {90} (\bibinfo {year} {2007})}\BibitemShut {NoStop}%
\bibitem [{\citenamefont {{Virtanen}}\ \emph {et~al.}(2020)\citenamefont
  {{Virtanen}}, \citenamefont {{Gommers}}, \citenamefont {{Oliphant}},
  \citenamefont {{Haberland}}, \citenamefont {{Reddy}}, \citenamefont
  {{Cournapeau}}, \citenamefont {{Burovski}}, \citenamefont {{Peterson}},
  \citenamefont {{Weckesser}}, \citenamefont {{Bright}} \emph
  {et~al.}}]{Virtanen_2020}%
  \BibitemOpen
  \bibfield  {author} {\bibinfo {author} {\bibfnamefont {P.}~\bibnamefont
  {{Virtanen}}}, \bibinfo {author} {\bibfnamefont {R.}~\bibnamefont
  {{Gommers}}}, \bibinfo {author} {\bibfnamefont {T.~E.}\ \bibnamefont
  {{Oliphant}}}, \bibinfo {author} {\bibfnamefont {M.}~\bibnamefont
  {{Haberland}}}, \bibinfo {author} {\bibfnamefont {T.}~\bibnamefont
  {{Reddy}}}, \bibinfo {author} {\bibfnamefont {D.}~\bibnamefont
  {{Cournapeau}}}, \bibinfo {author} {\bibfnamefont {E.}~\bibnamefont
  {{Burovski}}}, \bibinfo {author} {\bibfnamefont {P.}~\bibnamefont
  {{Peterson}}}, \bibinfo {author} {\bibfnamefont {W.}~\bibnamefont
  {{Weckesser}}}, \bibinfo {author} {\bibfnamefont {J.}~\bibnamefont
  {{Bright}}}, \emph {et~al.},\ }\bibfield  {title} {\bibinfo {title} {{SciPy
  1.0: Fundamental Algorithms for Scientific Computing in Python}},\ }\href
  {https://doi.org/https://doi.org/10.1038/s41592-019-0686-2} {\bibfield
  {journal} {\bibinfo  {journal} {Nat. Methods}\ }\textbf {\bibinfo {volume}
  {17}},\ \bibinfo {pages} {261} (\bibinfo {year} {2020})}\BibitemShut
  {NoStop}%
\bibitem [{\citenamefont {Van Der~Walt}\ \emph {et~al.}(2011)\citenamefont {Van
  Der~Walt}, \citenamefont {Colbert},\ and\ \citenamefont
  {Varoquaux}}]{van2011numpy}%
  \BibitemOpen
  \bibfield  {author} {\bibinfo {author} {\bibfnamefont {S.}~\bibnamefont {Van
  Der~Walt}}, \bibinfo {author} {\bibfnamefont {S.~C.}\ \bibnamefont
  {Colbert}},\ and\ \bibinfo {author} {\bibfnamefont {G.}~\bibnamefont
  {Varoquaux}},\ }\bibfield  {title} {\bibinfo {title} {The numpy array: a
  structure for efficient numerical computation},\ }\href
  {https://doi.org/10.1109/MCSE.2011.37} {\bibfield  {journal} {\bibinfo
  {journal} {Comp. Sci. Eng.}\ }\textbf {\bibinfo {volume} {13}},\ \bibinfo
  {pages} {22} (\bibinfo {year} {2011})}\BibitemShut {NoStop}%
\end{thebibliography}%

\end{document}